\documentclass[prd, twocolumn, nofootinbib, floatfix]{revtex4-1}
 
\usepackage{amsmath}
\usepackage{graphicx}
\usepackage{dcolumn}
\usepackage{bm}
\usepackage{epsfig}
\usepackage{amssymb,latexsym,mathrsfs}
\usepackage{graphicx}
\usepackage{color}
\usepackage{hyperref}

\usepackage{tabu}

\hypersetup{
    colorlinks=true,
    linkcolor=red,
    citecolor=blue,
} 

\newcommand{\be}{\begin{equation}}
\newcommand{\ee}{\end{equation}}
\newcommand{\bs}{\begin{split}} 
\newcommand{\bea}{\begin{eqnarray}}
\newcommand{\eea}{\end{eqnarray}}

\newcommand{\fnl}{f_{\rm NL}}

\begin{document}

\title{Designing an Inflation Galaxy Survey:\\{\it how to measure $\sigma(\fnl) \sim 1$
using scale-dependent galaxy bias}} 
\author{Roland de Putter, Olivier Dor\'{e}} 
\affiliation{Jet Propulsion Laboratory, California Institute of Technology, Pasadena, CA 91109\\
California Institute of Technology, Pasadena, CA 91125}

\begin{abstract}

The most promising method for measuring primordial non-Gaussianity
in the post-Planck era is to detect large-scale, scale-dependent galaxy bias.
Considering the information in the galaxy power spectrum,
we here derive the properties of a galaxy clustering survey that would optimize constraints
on primordial non-Gaussianity using this technique.
Specifically, we ask the question what survey design is needed to reach a precision
$\sigma(f_{\rm NL}^{\rm loc}) \approx 1$.
To answer this question, we calculate the sensitivity to $f_{\rm NL}^{\rm loc}$
as a function of galaxy number density, redshift accuracy and
sky coverage. We include the multitracer technique, which helps minimize cosmic variance noise,
by considering the
possibility of dividing the galaxy sample into stellar mass bins.
We show that the ideal survey for $f_{\rm NL}^{\rm loc}$ looks very different
than most galaxy redshift surveys scheduled for the near future. Since those are more or less
optimized for measuring the BAO scale, they typically require spectroscopic redshifts.
On the contrary, to optimize the $f_{\rm NL}^{\rm loc}$ measurement, a deep, wide, multi-band imaging survey is
preferred. An uncertainty $\sigma(f_{\rm NL}^{\rm loc}) = 1$ can be reached with a full-sky survey that is complete to an
$i$-band AB magnitude $i \approx 23$
and has a number density $\sim 8$ arcmin$^{-2}$. Requirements on the multi-band photometry are set by
a modest photo-$z$ accuracy $\sigma(z)/(1+z) < 0.1$ and the ability to measure stellar mass with a precision
$\sim 0.2$ dex or better (or another proxy for halo mass with equivalent scatter).
While here we focus on the information in the power spectrum,
even stronger constraints can potentially be obtained with the galaxy bispectrum.

\end{abstract}

\date{\today} 

\maketitle

\section{Introduction}

Measurements of the (non-)Gaussianity of the primordial density perturbations
provide a strong test of the physics of cosmic inflation.
The tightest current constraints come from the cosmic microwave background (CMB) temperature bispectrum.
In this paper we will focus on non-Gaussianity of the local type, which
is constrained by the Planck satellite to be $f_{\rm NL}^{\rm loc} = 2.7 \pm 5.8 \, (1\sigma)$ \cite{plancknongauss14}
(from here on we will drop the superscript ``loc'').

While the Planck data thus already limit any primordial non-Gaussianity to be very small,
important new insights into the nature of inflation may be gained by
probing non-Gaussianity with even higher sensitivity.
First of all,
in standard, single-field inflation, the level of primordial non-Gaussianity, as manifested in
the squeezed limit bispectrum, is suppressed to be unmeasurably small (for any experiment in the foreseeable future)
\cite{maldacena03,cremzal04}.
Thus, the current bounds leave a large range of values of $f_{\rm NL}$, which, if detected with a future survey,
would rule out all single-field models and thus prove that inflation involved multiple fields.
Moreover, multi-field models quite generically predict a non-Gaussianity level $|f_{\rm NL}| \gtrsim 1$
\cite{fnllsspaper,lythetal03,zal04}, not prohibitively far beyond current sensitivity.
Finally, various non-primordial effects, such as ``GR effects'' and non-linear evolution,
produce interesting cosmological signals equivalent to
those of primordial non-Gaussianity of order $|f_{\rm NL}| \sim 1$.

Partially based on the above, it would be extremely insightful to
improve our current bounds by a factor of a few to probe
$f_{\rm NL}$ of order unity. In this article, we will thus take $\sigma(f_{\rm NL}) \sim 1$ as an
approximate goal for future surveys.
Turning to the question of how to reach this goal,
we note that, unfortunately, the CMB is already close to its cosmic variance limited precision
and is not expected to reach beyond a precision $\sigma(f_{\rm NL}) \approx 2$ \cite{cmbpol09}, even when
optimal polarization data are included.
Fortunately, large-scale structure surveys, with their ability to probe three-dimensional
volumes and therefore large numbers of density modes, can in principle do better.
Specifically, primordial non-Gaussianity leads to a scale-dependent bias of halos,
and therefore of, e.g., galaxies, relative to the underlying matter density fluctuations
\cite{dalaletal08,matver08,slosaretal08,desjselj10}.
This effect has already been used to constrain $f_{\rm NL}$ at a precision
comparable to that reached by the WMAP CMB bispectrum analysis, $\sigma(f_{\rm NL}) \approx 20$
\cite{rossetal13,giannantonioetal14,leistedtetal14,wmap},
but it is possible to strongly improve on this with future galaxy surveys (recent studies include
\cite{ferramachoetal14,ferrsmith14,yamauchietal14,raccetal14,cameraetal14}).

The question we will address in this paper is then: {\it what would a galaxy survey need to look like
to reach $\sigma(f_{\rm NL}) \approx 1$?}
To obtain a clear connection to what is observed, we characterize
a survey directly in terms of the properties of the observed
{\it galaxy} sample. In particular, we describe the galaxy sample
in terms of a stellar mass cut taking into account scatter between stellar
mass and host halo mass (as opposed to describing the survey directly in terms of halo
properties, or in terms of a free number density and bias).
Using this approach, we first study the sensitivity to $f_{\rm NL}$
as a function of various survey properties, such as minimum stellar mass (i.e.~depth),
at various redshifts.

In a realistic survey, galaxy sample properties will vary with redshift.
Therefore, we will next use a simple model, based on a ($i$-band) magnitude cut, for this redshift evolution,
to estimate how the $f_{\rm NL}$ constraint builds up as a function of redshift.
We will then also compute $\sigma(f_{\rm NL})$ as a function of total survey depth,
defined by the magnitude cut, or equivalently by total number of galaxies per square degrees.

An important consideration in these studies is that, since the bias correction
due to $f_{\rm NL}$ is proportional to $k^{-2}$, where $k$ is the wave vector of a density mode in
Fourier space, the signal we are looking for is dominated by the largest scales accessible to a survey.
The constraining power is therefore strongly limited by cosmic variance. However, this cosmic variance
can partially be cancelled if multiple tracers, e.g.~multiple galaxy samples, with different biases are available
\cite{seljak09}.
We thus include this {\it multitracer} technique in our forecasts
and will discuss in detail the comparison between the multitracer and single-tracer approaches.

The paper is organized as follows. In Section \ref{sec:method},
we describe the effect of primordial non-Gaussianity on halo bias, the forecast method, and
the halo occupation distribution (HOD) prescription for populating halos with galaxies
of various stellar masses. Next, in Section \ref{sec:survey optimization}, we study the constraining power on $f_{\rm NL}$
as a function of survey depth, stellar mass scatter relative to the mean stellar mass-halo mass
relation, survey volume, and redshift accuracy. While in Section \ref{sec:survey optimization} we treat those survey properties
as free parameters at various redshifts, in Section \ref{sec:survey}, we will study a simple model
for the redshift dependence of these properties and quantify what type of survey is needed to
reach $\sigma(f_{\rm NL}) \sim 1$. We will also compare this to planned galaxy surveys.
Finally, we summarize and conclude in Section \ref{sec:conclusions}.

\section{Method}
\label{sec:method}

\subsection{Scale-dependent halo bias}

We will focus on primordial non-Gaussianity of the local type \cite{salopekbond90,komsper01}.
For this ansatz, it has been shown that the linear halo bias receives a scale-dependent correction
\cite{dalaletal08,matver08,slosaretal08,desjselj10},
\be
\label{eq:fnl bias}
b = b_G + 2 \, f_{\rm NL} \, (b_G - 1) \, \delta_c \, \frac{3 \Omega_m H_0^2}{2 k^2 T(k) D(z)}.
\ee
Here $b_G$ is the Eulerian, Gaussian halo bias and $\delta_c$ is the critical overdensity for
halo collapse, here taken to be the critical density for spherical collapse,
$\delta_c = 1.686$. Furthermore, $\Omega_m$ is the matter density at $z=0$ relative to
the critical density, $H_0$ the Hubble constant ($z=0$), $T(k)$ is the transfer function
of matter perturbations, normalized to $1$ at low $k$, and $D(z)$ is the linear
growth function, normalized such that $D(z) = 1/(1 + z)$ during matter domination\footnote{The
normalization choice for $D(z)$ implies that
$f_{\rm NL}$ is defined in terms of the Bardeen potential
at {\it early} times. Both the CMB literature and a large fraction
of the large-scale structure literature use this convention.
However, some works on large-scale structure define
$f_{\rm NL}$ in terms of the potential at $z=0$,
corresponding to normalizing $D(z=0)=1$ in Eq.~(\ref{eq:fnl bias}).
To add to the confusion, the former convention is sometimes referred to
as the ``CMB convention'' and the latter as the ``LSS convention''.
}.

A key feature of this bias correction that will be important later is that
the effect is proportional to $k^{-2}/T(k)$ and therefore most important on
large scales. The bias correction is of order $f_{\rm NL}$
at the horizon scale. Moreover, the bias correction is proportional to $b_G - 1$,
so that there is no scale dependence for an unbiased tracer.

\subsection{Fisher matrix forecasts}

We forecast constraints on $f_{\rm NL}$ from scale-dependent halo bias using
the Fisher matrix formalism (see e.g.~\cite{fisher}).
We consider the general case of multiple tracers with number densities
$\bar{n}_i$ and bias $b_{G,i}$.
Since the relevant information will come from large scales in the linear regime,
we will use the linear Kaiser model for the halo cross- and powerspectra,
\be
\label{eq:pk kaiser}
P_{ij}(k,\mu) = \left(b_i + f \mu^2 \right) \, \left(b_j + f \mu^2 \right) \, P_m(k) + \frac{1}{\bar{n}_i} \, \delta^K_{ij}.
\ee
Here, $\mu$ is the cosine of the angle between the wave vector ${\bf k}$
and the line of sight direction (the plane-parallel approximation is used),
$f = d\ln D/d\ln a$ is the linear growth rate and $P_m(k)$ is the linear matter power spectrum.
The bias $b_i$ for each species is given by Eq.~(\ref{eq:fnl bias}) in terms of the
bias $b_{G,i}$ in the absence of primordial non-Gaussianity.
Our Fisher forecasts assume Gaussianity of the halo overdensity itself to calculate the covariance
of the signal and we always assume a fiducial\footnote{At the level of precision
corresponding to $f_{\rm NL} \sim 1$, there are also relevant general relativistic terms
in the observed power spectrum that look like an effective $f_{\rm NL}$ of order unity, e.g.~\cite{JSH12}. Moreover,
some works suggest that the non-linearity of gravity induces an effective $f_{\rm NL} = -\frac{5}{3}$
in large-scale structure,
even in the case of single field inflation \cite{bartetalreview04,bartetallss05,verdemat09,brunietal14b,villaetal14}
(although see \cite{pajeretal13} for an opposing perspective).
We do not include these effects in this work and simply
use our fiducial $f_{\rm NL} = 0$ as an approximation to the expected signal in the single-field case.}
$f_{\rm NL} = 0$.

In Eq.~(\ref{eq:pk kaiser}), we have assumed that the stochastic component of the halo overdensity
is Poissonian. We show in Appendix \ref{sec:nonpoisson} that applying a more realistic, non-Poissonian
description based on the halo model, and corroborated by simulations \cite{hamausetal10}, has a small enough effect ($< 10 \%$ on the $\fnl$ uncertainty) on the results
that it can be ignored for our purposes.

Using Eq.~(21) of \cite{hamausetal12}, we can obtain a useful, simple expression
for the diagonal Fisher matrix component corresponding to $f_{\rm NL}$. This quantity
is equivalent to the (unmarginalized) signal-to-noise squared of $|f_{\rm NL}| = 1$
and given by
\bea
\label{eq:fm}
F_{f_{\rm NL} f_{\rm NL}} &=&  \left[ \Sigma^{-1} \left( \Sigma_{f_{\rm NL} f_{\rm NL}}
 + \frac{\Sigma^2_{f_{\rm NL}}}{\Sigma} \right) + \Sigma_{f_{\rm NL} f_{\rm NL}} - \frac{\Sigma^2_{f_{\rm NL}}}{\Sigma}\right] \nonumber \\
&\times& \left( 1 + \Sigma^{-1} \right)^{-2}
\eea
with
\bea
\Sigma &\equiv& \sum_i \bar{n}_i \, b_i^2  \nonumber \\
\Sigma_{f_{\rm NL}} &\equiv& \sum_i \bar{n}_i \, b_i \, \frac{\partial b_i}{\partial \fnl} \nonumber \\
\Sigma_{f_{\rm NL} \fnl} &\equiv& \sum_i \bar{n}_i \, \left(\frac{\partial b_i}{\partial \fnl}\right)^2 \nonumber.
\eea
We refer to \cite{hamausetal12} for more details on the derivation of this expression.
Following said reference, the first and second terms in the square brackets
in Eq.~(\ref{eq:fm}) can
be identified as single- and multitracer contributions respectively.
Indeed, in the case of a single tracer, with $\bar{n}_i = \bar{n}$
and $b_{i} =  b$, the second term vanishes and the Fisher information
becomes
\be
F_{\fnl \fnl} = 2 \, \left(\frac{\bar{n} b^2}{1 + \bar{n} b^2}\right)^2 \, \left(\frac{\partial \ln b}{\partial \fnl}\right)^2.
\ee
In the limit of no shot noise ($\bar{n} \to \infty$), this reduces to
$F_{\fnl \fnl} = 2 (\partial \ln b/\partial \fnl)^2$. The Fisher information
in the single-tracer case thus has an upper limit due to cosmic variance.
In the multitracer case, this cosmic variance can be evaded.
Indeed, the multitracer term is unbounded in the limit of no shot noise
($\bar{n}_i \to \infty$), see e.g.~\cite{seljak09,hamausetal12}.

Projected uncertainties on $\fnl$ are in principle obtained by
calculating the Fisher matrix for all cosmological and nuisance parameters
and marginalizing over them. However, as demonstrated in Appendix \ref{sec:marg},
marginalization has only a modest effect on $\sigma(\fnl)$, increasing it by
at most $\sim 40 \%$. For simplicity, we therefore in the body of this paper focus on unmarginalized
constraints, i.e.
\be
\sigma(\fnl) = 1/\sqrt{F_{\fnl \fnl}}.
\ee

The $\fnl$ constraint depends strongly on the minimum wave vector included in the analysis, and,
to a lesser extent on the maximum wave vector. Our default choices are $k_{\rm min} = 0.001 h/$Mpc
and $k_{\rm max} = 0.1 h/$Mpc. We investigate the dependence on the range of scales available
in Sections \ref{subsec:vol} and \ref{subsec:sigz}. In particular, the largest included scale
is approximately given by $k_{\rm min} = \pi/V^{1/3}$, where $V$ is the survey volume, so
that our default choice $k_{\rm min} = 0.001 h/$Mpc corresponds to a survey of $V \sim 30 \, (h^{-1} $Gpc$)^3$.

\subsection{From halos to galaxies}
\label{subsec:hod}

The previous subsection explained how to forecast $\fnl$ constraints
in terms of
the number densities and biases of a set of halo (sub)samples.
In this paper we study the constraints from galaxy clustering and
we would like to define the galaxy sample(s)
in terms of an observable galaxy property.
We choose this to be the stellar mass $M_*$, which is related
to the galaxy's intrinsic brightness, and, in more detail,
depends on a galaxy's initial mass function,
stellar evolution, dust content and metallicity of its stars.

Then, for galaxy samples defined by stellar mass cuts, we compute the corresponding
number densities and biases of host halos using the following approach.
We start from the description of the halo number density and bias as a function
of {\it halo mass}, calibrated on simulations and given by, respectively, \cite{tinkeretal08} and \cite{tinkeretal10}.
We then insert a central galaxy into each halo,
with a stellar mass drawn from the halo occupation distribution (HOD)
prescription of \cite{L12,nipotietal12}.
Specifically, for a given halo mass $M_h$, the expectation
value of the logarithm of the stellar mass of the central galaxy, $\langle \log_{10} M_* \rangle \equiv \log_{10} M_{*,SHMR}$
(with SHMR standing for stellar-to-halo mass relation), is implicitly defined by the relation
\bea
\label{eq:stel2hal}
\log_{10}(M_h) &=& \log_{10}(M_1) + \beta \log_{10}\left( \frac{ M_{*,SHMR} }{M_{*,0}}\right) \nonumber \\
&+& \frac{\left( M_{*,SHMR}/M_{*,0}\right)^\delta}{1 + \left(M_{*,SHMR}/M_{*,0}\right)^{-\gamma}} - \frac{1}{2}
\eea
(this relation and its parameter dependence is nicely illustrated in Fig.~3 of \cite{L12}).
The distribution of $M_*$ around this mean is taken to be a Gaussian in $\log_{10}(M_*)$
with standard deviation $\sigma_{\log M_*} \equiv \sigma(\log_{10} M_*)$.
At $z = 0.22 - 1$, we use the values of the parameters
$M_1$, $M_{*,0}$, $\delta$, $\gamma$ and $\sigma_{\log M_*}$ from Table 5 (SIGMOD1)
of \cite{L12}. These are based on a fit to abundance, clustering and galaxy-galaxy
lensing data.
Stellar mass in these works is obtained by fitting ground-based COSMOS photometry in eight bands
to a grid of models for the galaxy's spectral energy distribution (SED).
To get the redshift dependence of the HOD parameters outside
of the range $z=0.22- 1$, we follow {\it prescription (iii)}
of \cite{nipotietal12} \, (Sec 3.2.1), which makes use of fitting formulas linear in scale factor,
given in \cite{B10}.

Since we make heavy use of the HOD parameters from \cite{L12},
we adopt the same fiducial cosmology, namely a WMAP5, spatially flat $\Lambda CDM$
model:
$\Omega_m = 0.258$, $\Omega_b h^2 = 0.02273$, $n_s = 0.963$,
$\sigma_8 = 0.796$, $h = 0.72$, and $\fnl = 0$.

\begin{figure*}[htbp!] 
\includegraphics[width=0.48\textwidth]{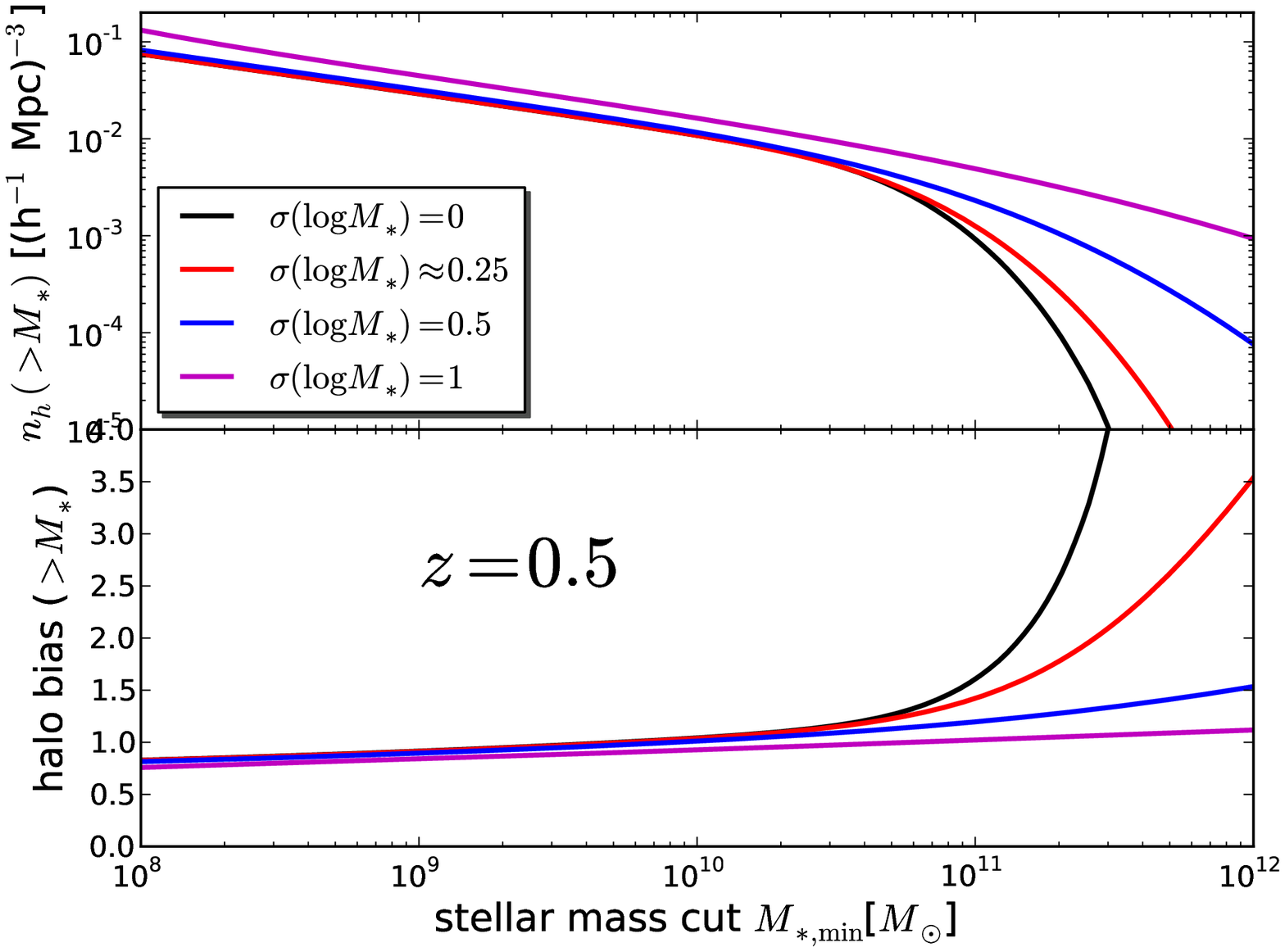}
\includegraphics[width=0.48\textwidth]{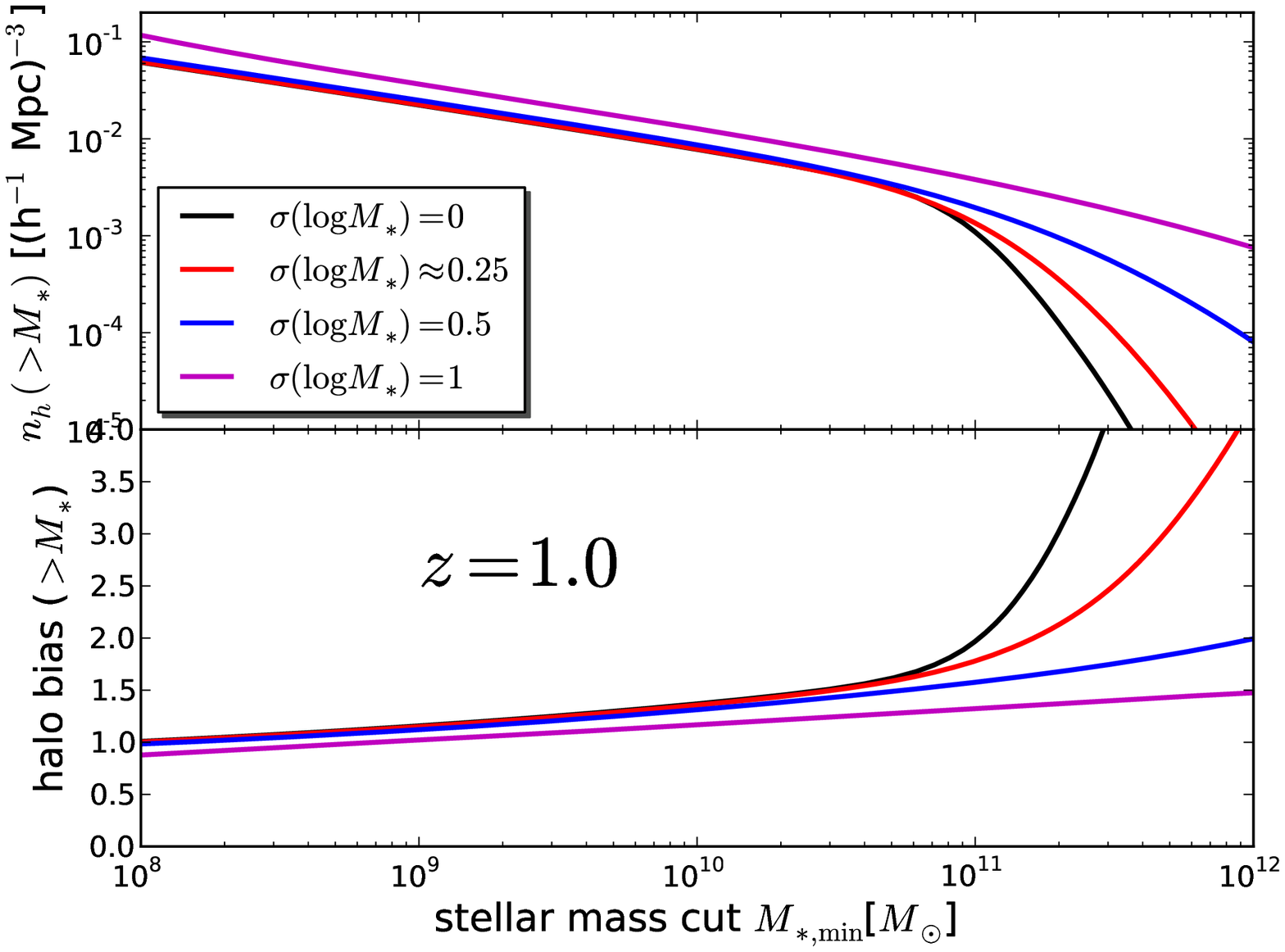}
\includegraphics[width=0.48\textwidth]{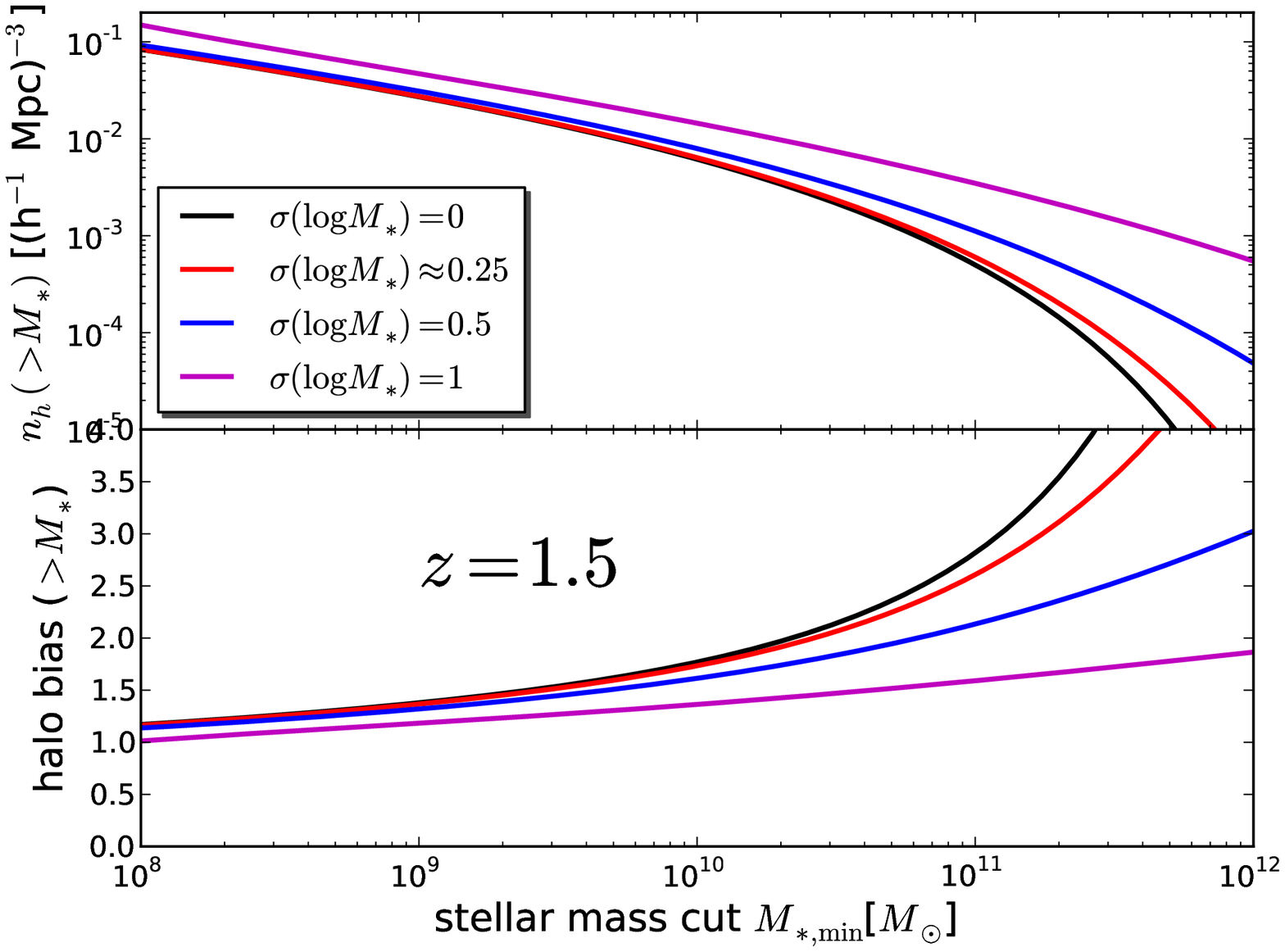}
\includegraphics[width=0.48\textwidth]{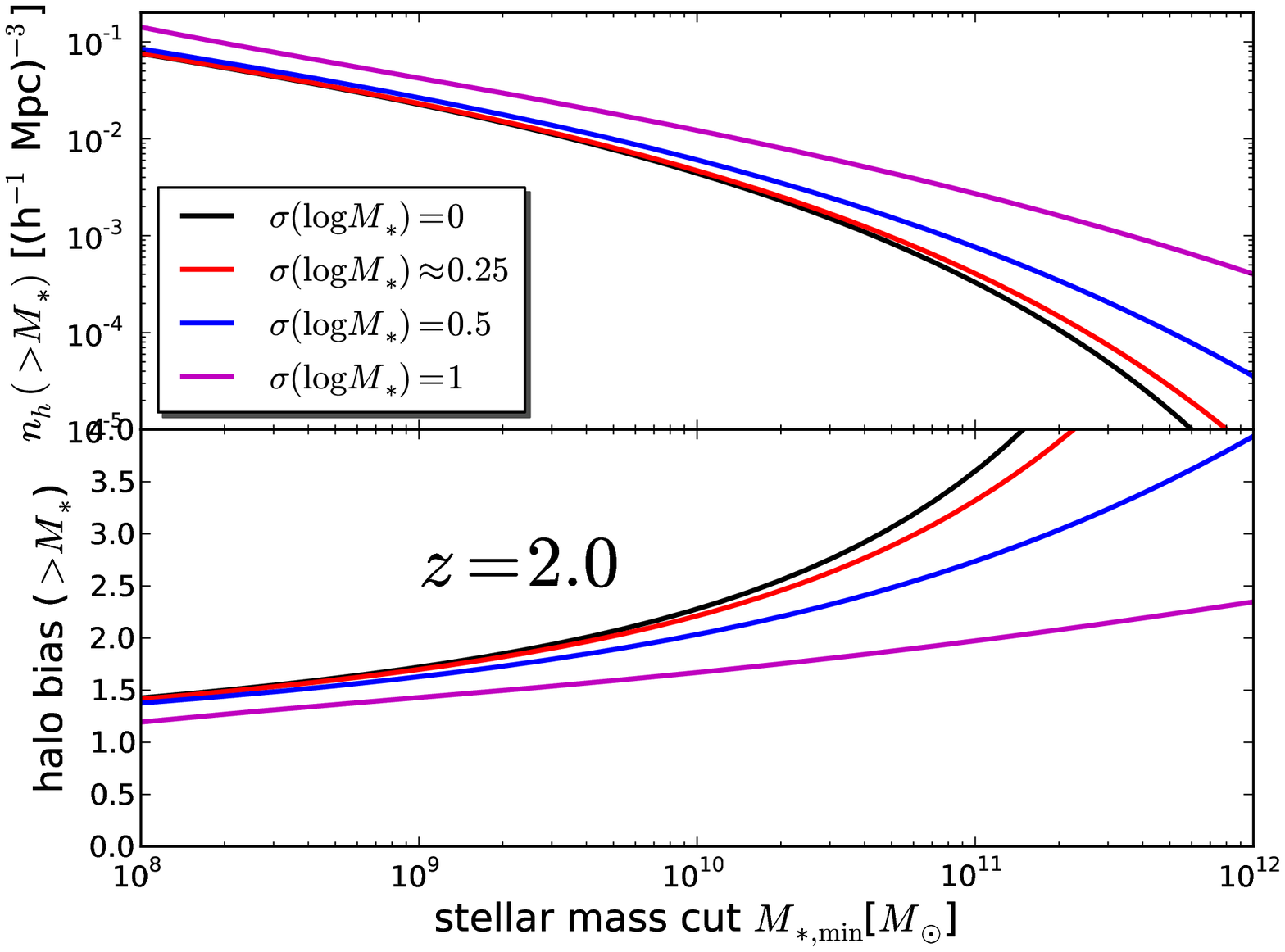}
\caption{The number density and bias of galaxies above a stellar mass cut $M_*$ at
various redshifts.
Different curves are for different assumed scatter in
stellar mass, $\sigma_{\log M_*}$ (see text),
relative to a deterministic stellar mass-halo mass relation, Eq.~(\ref{eq:stel2hal}).
The red curves correspond to the fiducial model for $\sigma_{\log M_*}$,
based on simulations.
The number density in the top panel
is the number densities of central galaxies, or, equivalently, of different halos
and the bias is the mean bias of those halos.
Brighter objects typically live in more massive halos, which are rarer, and therefore have a larger bias.
}
\label{fig:num and bias vs mstar}
\end{figure*}

We will characterize our galaxy sample at a given redshift by a simple stellar mass
cut, $M_* > M_{*,{\rm min}}$, and equate central galaxies with halos (we will come back to this in the last paragraph
of this section).
The red curves in Figure \ref{fig:num and bias vs mstar} show the resulting number density
of halos, or equivalently, of central galaxies, and the mean bias of those halos,
as a function of $M_{*,{\rm min}}$. For the single-tracer forecasts, we will use the number density
and bias plotted. In the multitracer case, we divide the sample $M > M_{*,{\rm min}}$
into a large number of stellar mass bins. We make sure to use a large enough number of bins
that the constraining power has converged, i.e.~we show the optimal multitracer constraints
possible with a stellar mass binning. In Appendix \ref{sec:mass bins}, we explore how many
bins are needed and find that typically two or three bins is already close to optimal
(see \cite{hamausetal12} for a study of optimal weighting schemes of halo samples).

The red curves in Figure \ref{fig:num and bias vs mstar} show the
fiducial case from \cite{L12,nipotietal12}, where $\sigma_{\log M_*} \approx 0.25$.
This scatter includes both the intrinsic scatter between halo mass and its
corresponding stellar mass, and the scatter due to measurement uncertainty,
the latter of course being specific to the COSMOS sample used as input in \cite{L12}.
We will take this as our fiducial model, but will
consider various values for $\sigma_{\log M_*}$, particularly in Section \ref{subsec:sigm}.
These values can represent differing levels of noise in the stellar
mass determination, but also, crudely, the use of a different proxy for halo mass,
such as a flux in a certain wavelength band, that
may have a different scatter than stellar mass.
The various colors in Figure \ref{fig:num and bias vs mstar} show the effect on
number density and bias of modifying $\sigma_{\log M_*}$.
We will come back to this in Section \ref{subsec:sigm}.

In our approach, we implicitly equate positions of {\it central} galaxies with the positions
of their host halos and we do not make use of satellite galaxy positions since they live in
halos for which we already have a central galaxy.
In other words, we equate central galaxy number densities with the halo number densities
that enter our Fisher matrix calculation.
In practice, there are of course complications to this picture. First of all, the central galaxy
does not perfectly match the center of its host halo, and secondly, it is not always possible to
separate centrals from satellites. In reality, it is thus more practical to simply use all galaxies. This corresponds
to a reweighting of halos and thus affects the bias of each sample.
However, for the sake of forecasts of the approximate information
content of future surveys, our approach should be sufficient.
The {\it total} number density of observed galaxies, often used in this paper, of course does include both central
and satellite galaxies. Based again on the HOD study in \cite{L12}, we find that
the satellite galaxy fraction is relatively independent of stellar mass and of redshift,
$f_{\rm sat} \approx 0.25$. For simplicity, we thus use $f_{\rm sat} = 0.25$
to relate halo/central number density to {\it total} galaxy number density.

\section{Dependence of sensitivity to $\fnl$ on survey properties}
\label{sec:survey optimization}

\begin{figure*}[htbp!]
\includegraphics[width=0.48\textwidth]{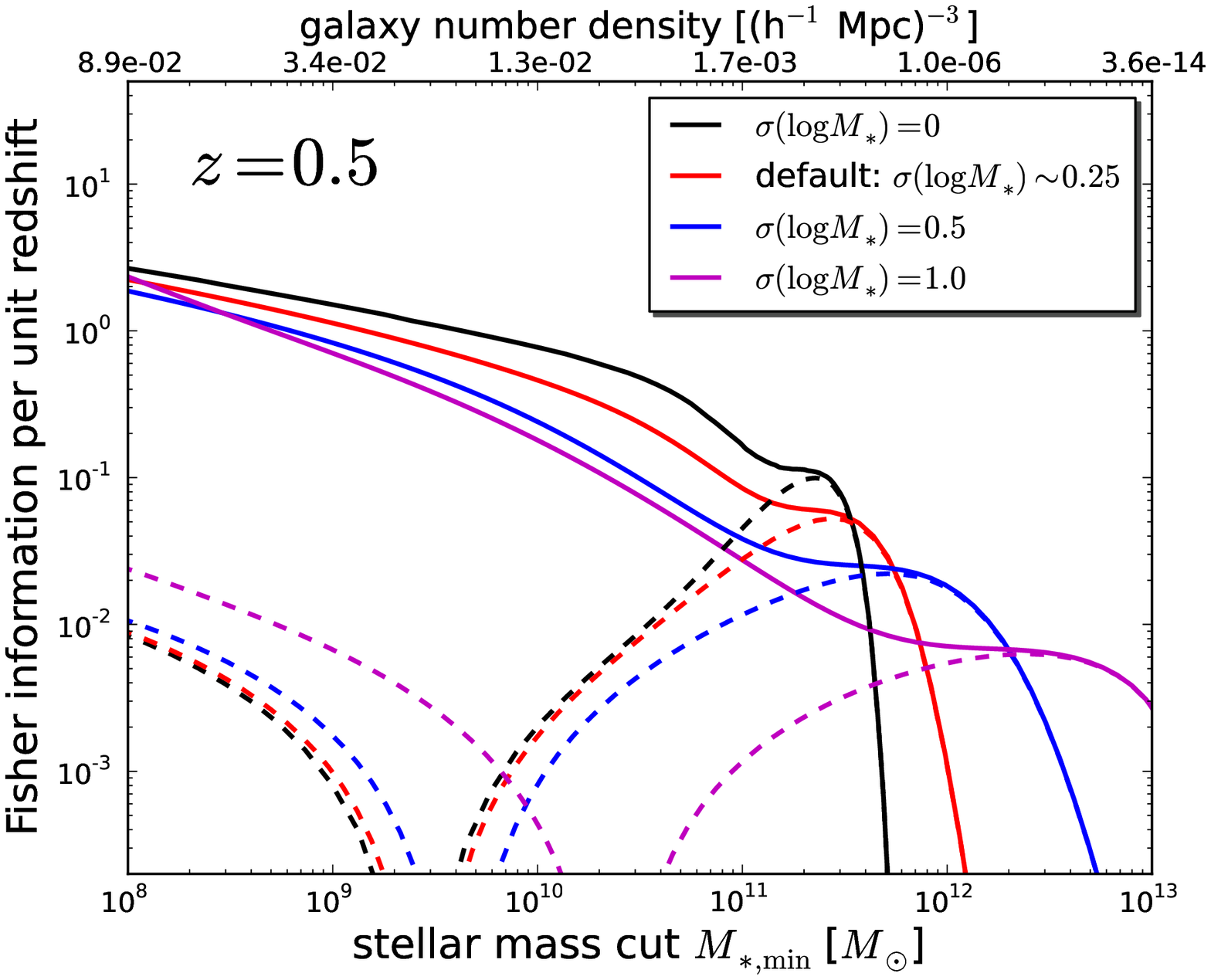}
\includegraphics[width=0.48\textwidth]{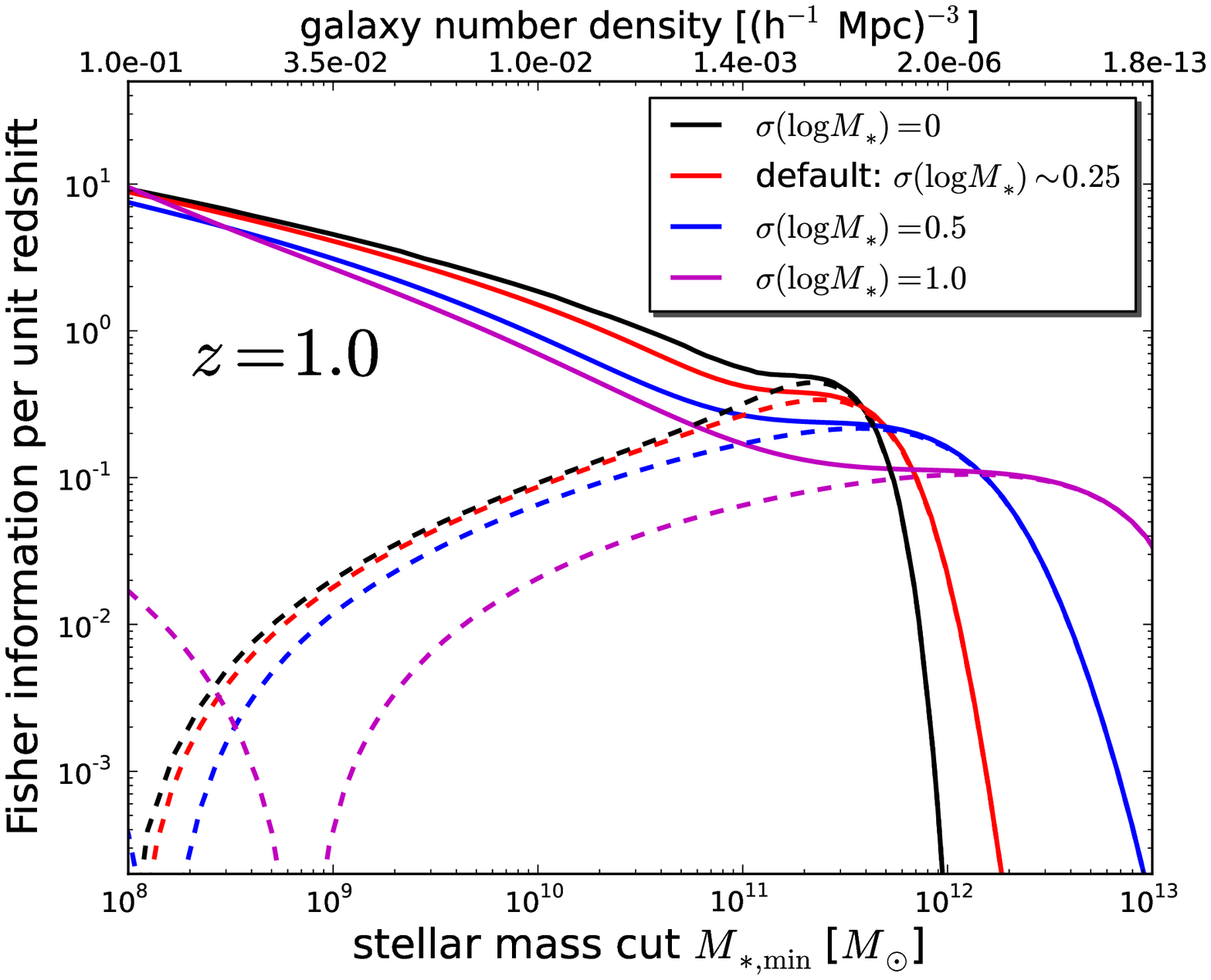}
\includegraphics[width=0.48\textwidth]{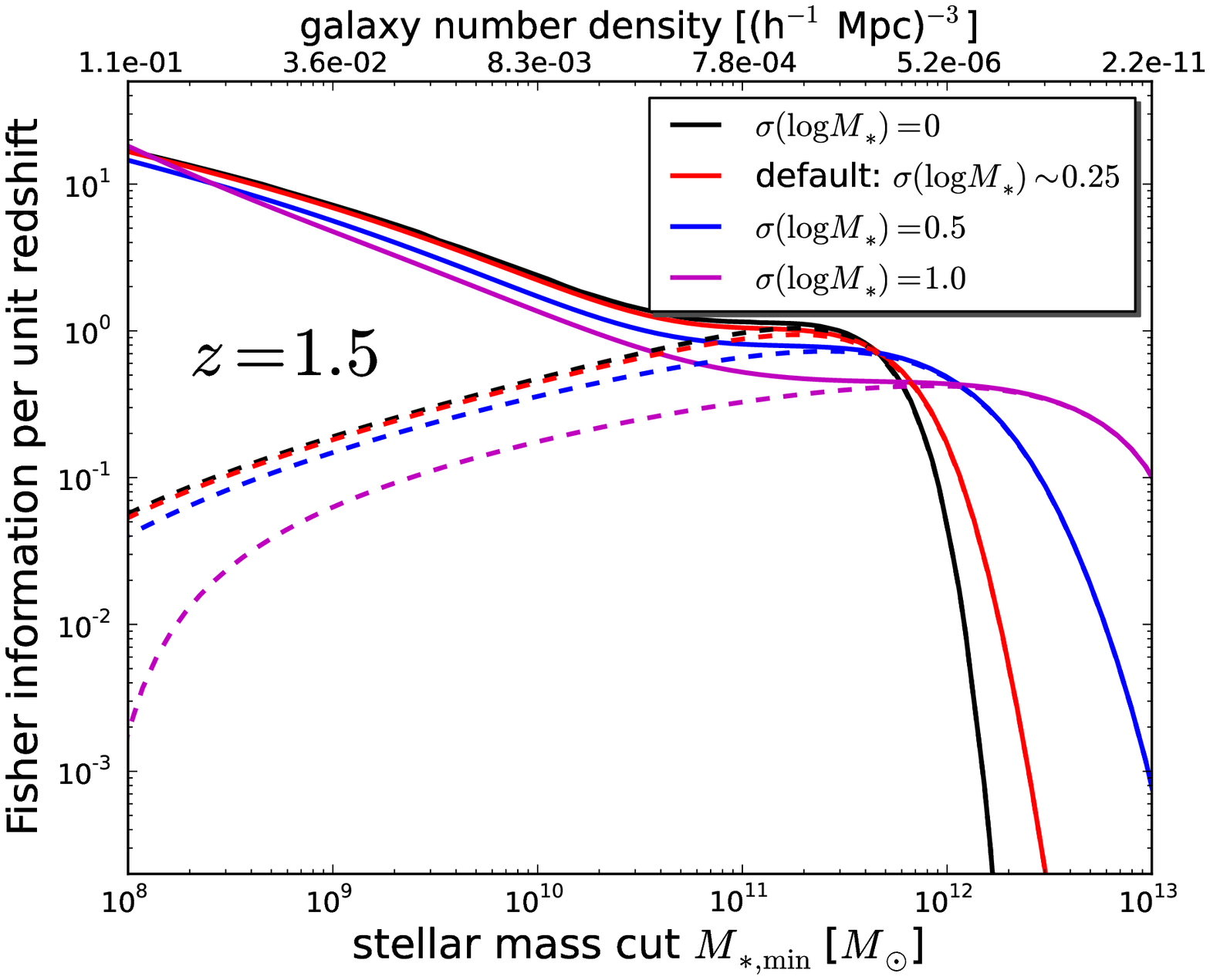}
\includegraphics[width=0.48\textwidth]{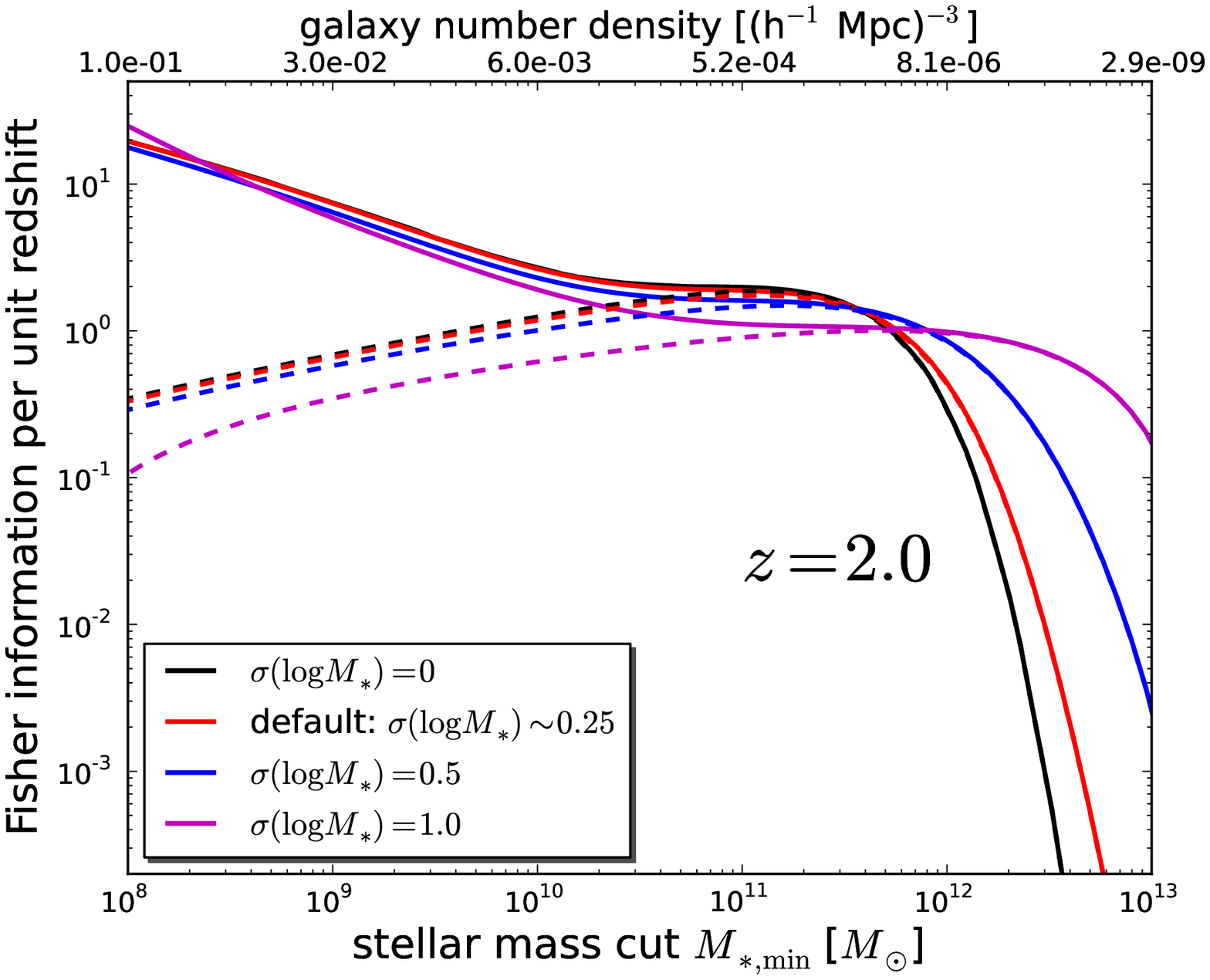}
\caption{The differential Fisher information in $f_{\rm NL}$ per unit redshift as a function
of minimum {\it stellar mass}, assuming the {\it galaxy} sample consists of all galaxies above the stellar mass cut.
We assume a sky coverage $f_{\rm sky} = 0.75$.
Dashed lines show the single-tracer case and solid lines show the optimal multitracer
case. Different colors correspond to different assumed scatters in $\log_{10}M_*$
at fixed host halo mass, where $M_*$ is the stellar mass of the halo's central galaxy.
The number density on the upper horizontal axis is the total galaxy number density (central plus satellite)
corresponding to the minimum stellar mass $M_*$ on the lower axis, in the case of the default
$\sigma(\log M_*) \approx 0.25$.
The information on $\fnl$ goes through zero when the mean halo bias of the sample equals unity. For instance, at $z = 0.5$,
this occurs at a stellar mass cut $M_{*,{\rm min}} \approx 4 \cdot 10^{9} M_\odot$ (cf.~the bias in Figure \ref{fig:num and bias vs mstar}),
with the exact value depending on the assumed stellar mass scatter.}
\label{fig:FMgal}
\end{figure*}

We now study $f_{\rm NL}$ constraining power at various fixed redshifts
and its dependence on survey properties.
We will pay particular attention to the comparison between the single- and multitracer
case.
At the end of this section, we will use the various dependencies to draw conclusions about
what an ideal $f_{\rm NL}$ survey may look like.
In the next section, we will then consider a toy model for such a survey, taking
into account that it may cover a wide redshift range and
that survey properties like $\bar{n}_i$ and $b_i$
may vary with redshift.

Since in this section, we focus on the constraining power at fixed redshifts,
we will often quantify it in terms of the Fisher information
per unit redshift for a ``full-sky'' ($f_{\rm sky} = 0.75$) survey, $dF_{\fnl \fnl}/dz$.
This is equivalent to the signal-to-noise squared per unit redshift for a signal
$|\fnl = 1|$. It is a useful quantity because, unlike $\sigma(\fnl)$,
it is additive when combining different redshifts. Keep in mind, however,
that the uncertainty on $\fnl$ scales like one divided by the {\it square root}
of the total Fisher matrix (i.e.~$dF_{\fnl \fnl}/dz$ integrated over redshift), so that variations in $\sigma(\fnl)$ are less dramatic than those
in the Fisher information.

\subsection{Survey depth (minimum stellar mass $M_{*,{\rm min}}$)}
\label{subsec:mmin}

The red curves in Figure \ref{fig:FMgal} show the Fisher information per unit redshift
as a function of $M_{*,{\rm min}}$ for the fiducial stellar mass scatter ($\sigma_{\log M_*} \sim 0.25$).
We assume a sky coverage $f_{\rm sky} = 0.75$.
The dashed curves depict the single-tracer case and the solid curves are for the
multitracer scenario. The horizontal axis on top gives the corresponding comoving
number density of galaxies, both central and satellites,
but for a more detailed look at the $M_{*,{\rm min}}$ - number density relation,
we refer to Figure \ref{fig:num and bias vs mstar}.
Let us first consider the $z=0.5$ single-tracer case (dashed).
If we start at low number density/high $M_{*,{\rm min}}$ on the right of the plot,
and move towards the left,
the clustering measurement is first shot noise dominated
and the constraining power improves rapidly as $M_{*,{\rm min}}$ is lowered
and the number density increased.
At some point, when $M_{*,{\rm min}} \sim 2 \cdot 10^{11} M_\odot$, the improvement
becomes weaker and the constraining power even starts decreasing. There are two reasons for this behavior.
On the one hand, as the number density gets high enough for the clustering measurement to become
sample variance dominated, there are, for fixed signal, no gains from improving the number density further.
On top of this, as $M_{*,{\rm min}}$ is lowered, the bias starts to approach unity, and since the $\fnl$
signature is proportional to $b_G - 1$, the signal weakens. This explains that the curve does not just
reach a plateau, but in fact curves down. Going further to the left, the constraining power vanishes
when the mean bias equals one and after that starts increasing again when the bias drops below one.

It is well known that the multitracer technique only adds information
in the high number density regime (e.g.~\cite{seljak09,hamausetal11}) and this is indeed what the solid curve shows.
If again we move from right to left in the $z=0.5$ plot, we find that for low number density,
the multitracer and single-tracer cases have the same constraining power.
Then, when the single-tracer curve starts to turn over, we find that first the multitracer curve remains
constant on a small plateau. While the information from the multitracer forecast is thus larger here than that from the single-tracer
case with the same $M_{*,{\rm min}}$, it is still not the use of multiple tracers that adds information: 
the multitracer constraint is simply equivalent to the single-tracer constraint with a larger $M_{*,{\rm min}}$,
i.e.~it is like throwing away the low-$M_*$ end of the distribution in order to prevent the mean bias to approach one.
However, pretty soon the multitracer curve does indeed go {\it up} and this is where the use of multiple tracers
starts to pay dividends. Significant gains are reached for comoving number densities $\bar{n} > 10^{-3} (h^{-1}$Mpc$)^{-3}$
and $M_{*,{\rm min}} < 5 \cdot 10^{10} M_\odot$. In principle the multitracer technique can boost the information by several
orders of magnitude. However, very large number densities and low stellar mass objects are required.
Another interesting approach for boosting the $\fnl$ signal that we do not consider here is to apply a more optimal
weighting scheme to each sample \cite{hamausetal10}. This could in principle improve our single-tracer constraints.
However, out multitracer approach is already optimal as it subdivides the sample into a large number of subsamples,
from which the Fisher matrix formalism subsequently extracts all information.

At higher redshift, the comparison between single- and multitracer is qualitatively similar,
showing the need for deep samples in order to take advantage of the multitracer benefits.
A major change at higher redshifts is that
for a given stellar mass cut, the bias is larger and therefore the point where
the single-tracer analysis loses its power ($b_G \to 1$) is shifted to much lower $M_{*,{\rm min}}$.

Looking next at the absolute level at which $\fnl$ could be constrained,
first note that the target level $\sigma(\fnl) \sim 1$ roughly corresponds to
$d$FM$/dz \sim 1$ for a redshift interval of order unity.
Figure \ref{fig:FMgal} thus shows that a single-tracer survey with $M_{*,{\rm min}} \approx 2 \cdot 10^{11} M_\odot$
out to $z \sim 2$ could marginally reach $\sigma(\fnl) \sim 1$. A deeper survey with $M_{*,{\rm min}} \approx 5 \cdot 10^{10} M_\odot$
or smaller would unlock the power of the multitracer technique and would reach stronger $f_{\rm NL}$ bounds per unit volume.
We will come back in more quantitative detail to the question of the $\fnl$ uncertainty for different survey designs
in Section \ref{sec:survey}.

Finally, we note that Figure \ref{fig:FMgal} is consistent with Figures 10-12 of \cite{hamausetal11},
which show the same calculation, but in terms of halo mass instead of stellar mass. At low redshift ($z \lesssim 1$),
the multitracer approach starts to pay off at $M_{h,{\rm min}} \sim 1 - $few $\times 10^{12} M_\odot$,
which indeed agrees with stellar mass $5 \cdot 10^{10} M_\odot$ or better.
We remind the reader that this corresponds to rather large comoving number densities, $\bar{n} \gg 10^{-3} (h^{-1} $Mpc$)^{-3}$.

\subsection{Stellar mass scatter $\sigma_{\log M_*}$}
\label{subsec:sigm}

Next we consider the effect of the scatter in $\log_{10} M_*$,
also shown in Figure \ref{fig:FMgal}.
Note that the relation between $M_{*,{\rm min}}$ and number density
depends on $\sigma_{\log M_*}$ and that the number densities on the top horizontal
axis are only valid in the fiducial model ($\sigma_{\log M_*} \sim 0.25$).
While the intrinsic scatter between halo mass and stellar mass
is fixed, variations in $\sigma_{\log M_*}$ represent the use of a different proxy for
mass than $M_*$ itself and/or the effect of measurement uncertainty in $M_*$ (although a constant
log scatter would not be the most natural choice to model measurement errors).
In addition to the default $\sigma_{\log M_*} \approx 0.25$ case, we show $\sigma_{\log M_*} = 0,0.5$ and $1.0$,
where the $\sigma_{\log M_*} = 0$ case is equivalent to having a zero-scatter proxy for halo mass.

For a fixed $M_{*,{\rm min}}$, a change in $\sigma_{\log M_*}$ affects the number density and bias of the sample
as well as the number densities and biases of the subsamples in the multitracer case.
As shown in Figure \ref{fig:num and bias vs mstar},
an increase in scatter (Eddington bias) leads to larger number density due to the negative curvature of
the mass function, and, consequently, a decrease in bias. Even if $M_{*,{\rm min}}$ is adjusted to keep the number density
fixed, the bias still decreases. A lower bias (for $b_G > 1$) leads to a smaller signal from $\fnl$
and, in the multitracer case, increasing $\sigma_{\log M_*}$ additionally leads to
smaller differences between the biases of the subsamples, which is detrimental.

Indeed, the horizontal shifts seen in Figure \ref{fig:FMgal} between different scatters,
are explained by the fact that, for larger $\sigma_{\log M_*}$, the same number density can be achieved
with a larger $M_{*,{\rm min}}$. The vertical shift, giving a decrease in information for larger
stellar mass scatter, is explained by the fact that, even for fixed number density, the halo bias goes down.
Comparing for example the constraints at the peaks of the single-tracer
curves (i.e.~the optimal constraint possible from single-tracer),
the effect is quite strong. Depending on redshift, a scatter $\sigma_{\log M_*} = 1$ gives a constraining power
that is a factor $\sim 2$ ($z=2$) to $\sim 10$ ($z=0.5$) weaker than in the default scenario (at low redshift,
a large stellar mass scatter has a larger effect on the sample
and leads to a  stronger reduction in mean bias).
Thus, it is paramount to measure a galaxy property that has as small a scatter relative
to halo mass as possible.

\subsection{Survey volume}
\label{subsec:vol}

\begin{figure*}[htbp!] 
\includegraphics[width=0.48\textwidth]{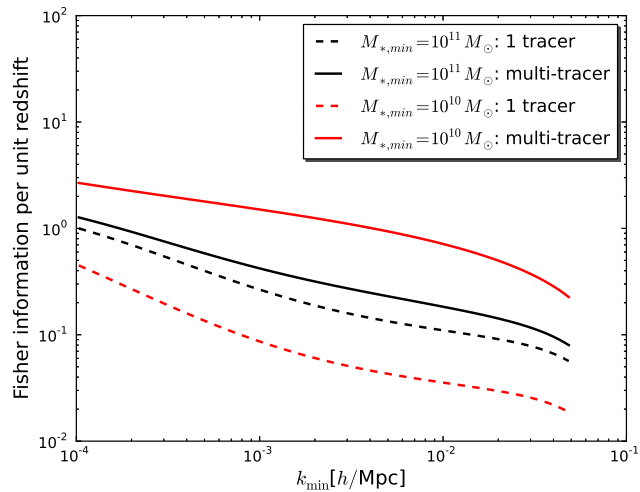}
\caption{Fisher information per unit redshift at $z = 1$ as a function of the minimum
wave vector $k_{\rm min}$ included in the Fisher matrix (sky coverage $f_{\rm sky} = 0.75$).
We fix $k_{\rm max} = 0.1 h/$Mpc. We show results
for a moderate density survey (black), $M_{*,{\rm min}} = 10^{11} M_\odot$, for which the multitracer
technique (solid) improves the $\fnl$ constraint little over the single-tracer case (dashed), but also
for a very dense survey (red), $M_{*,{\rm min}} = 10^{10} M_\odot$, for which using multiple tracers
improves the signal-to-noise squared by an order of magnitude.
In all cases, the constraining power strongly improves
as $k_{\rm min}$ is lowered.}
\label{fig:FM vs kmin}
\end{figure*}

\begin{figure*}[htbp!]
\includegraphics[width=0.48\textwidth]{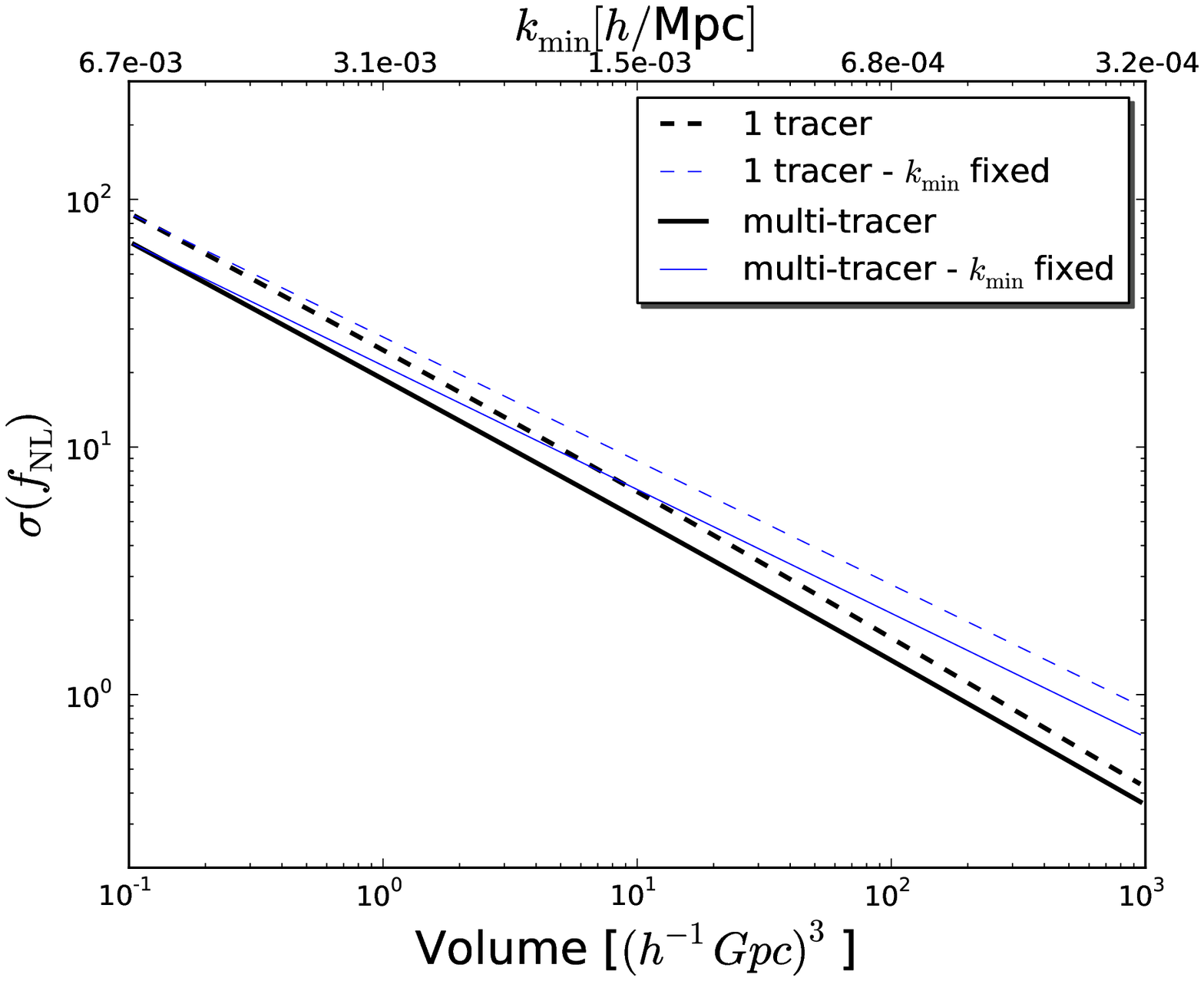} 
\includegraphics[width=0.48\textwidth]{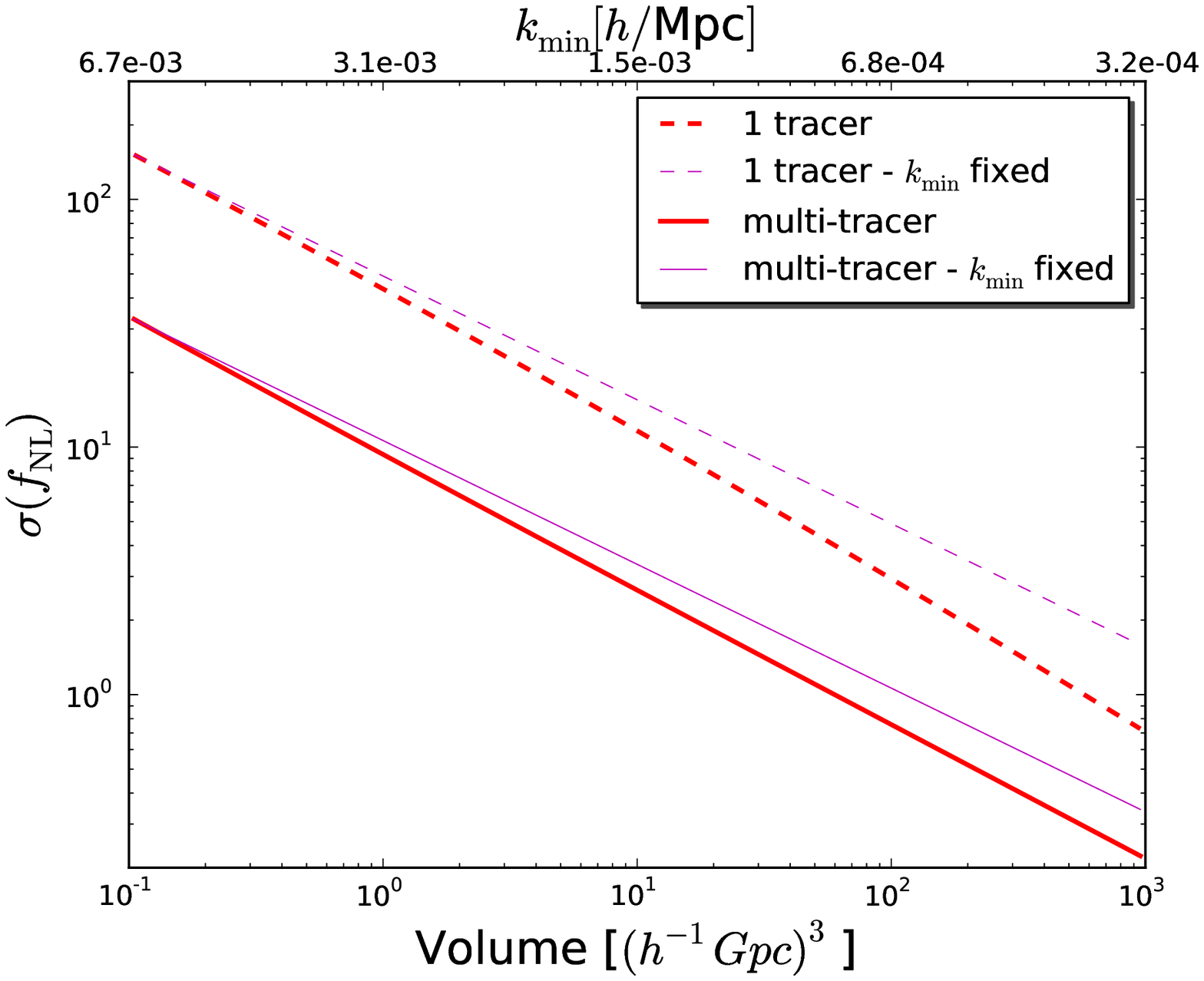} 
\caption{Uncertainty in $f_{\rm nl}$ as a function of comoving survey volume $V$ based on the Fisher information
per unit volume calculated at $z=1$.
Thick curves take into account the variation with survey volume of the minimum wave vector that can be used in the
analysis $k_{\rm min} = \pi/V^{1/3}$. Thin curves fix $k_{\rm min}$ to the value corresponding to the smallest volume
shown, $V = 0.1 (h^{-1}$Gpc$)^3$. {\it Left}: moderate density survey, $M_{*,{\rm min}} = 10^{11} M_\odot$.
{\it Right:} high density survey, $M_{*,{\rm min}} = 10^{10} M_\odot$.
An important advantage of large volume surveys is the ability to measure very
large modes (small $k_{\rm min}$).
In order to constrain primordial non-Gaussianity
at the level $\sigma(\fnl) \sim 1$, it is crucial to use a very large volume $V > 100 (h^{-1} $Gpc$)^{-3}$.}
\label{fig:FM vs vol}
\end{figure*}

Now that we have established
the dependence on galaxy sample and on the accuracy of mass discrimination,
we next study the dependence of expected $\fnl$ bounds on
survey volume $V$.
First of all, increasing the volume increases the number of galaxy overdensity modes
available within a fixed volume in {\bf k} space. This on its own leads to a $\propto 1/\sqrt{V}$
scaling of $\sigma(\fnl)$. On top of this, a larger volume allows us to probe clustering
on larger scales, i.e.~to use a smaller $k_{\rm min}$.
We illustrate the importance of the largest scales included in the Fisher matrix
in Figure \ref{fig:FM vs kmin}. The figure shows the Fisher information
per unit redshift at $z=1$ (again, assuming $f_{\rm sky} = 0.75$) as a function of
$k_{\rm min}$ for fixed $k_{\rm max}$. As a reminder, the default value used in this article
is $k_{\rm min} = 0.001 h/$Mpc.
We again show the single-tracer case in dashed and multitracer in solid. The different colors
represent different survey depths, $M_{*,{\rm min}} = 10^{11}$ and $10^{10} M_\odot$ (default $\sigma_{\log M_*}$).

Figure \ref{fig:FM vs kmin} displays a strong $k_{\rm min}$ dependence. This is expected
as, at low $k$, $F_{\fnl \fnl} \sim \int d\ln k k^{-1}$, which has an infrared divergence.
This is an additional reason to push for large survey volume (smaller $k_{\rm min}$).
In Figure \ref{fig:FM vs vol}, we next consider directly the dependence of $\sigma(\fnl)$
on survey volume by approximating $k_{\rm min} = \pi/V^{1/3}$ (thick curves).
We assume a survey centered at $z=1$ and do not take into account redshift evolution within the survey
volume (we include redshift evolution in Section \ref{sec:survey}).
To highlight the importance of the variation in $k_{\rm min}$,
the thin lines show the $\fnl$ uncertainty for the case of {\it fixed}
$k_{\rm min} = 6.7 \cdot 10^{-3} h/$Mpc, corresponding to a volume $V = 0.1 (h^{-1}$Gpc$)^{-3}$ (the smallest volume
included in the plot).

Compared with, e.g., information on baryon acoustic oscillations and most cosmological parameters,
the decrease of $k_{\rm min}$
with increasing $V$ is a much more important effect for $\fnl$, adding extra information compared to what would be
expected based on a simple $\propto 1/\sqrt{V}$ scaling.
Figure \ref{fig:FM vs vol} shows that an order unity uncertainty on $\fnl$
can be achieved with a moderate density (essentially single-tracer) survey, $M_{*,{\rm min}} = 10^{11} M_\odot$,
of volume $V \gtrsim 300 (h^{-1}$Gpc$)^{-3}$, or with a very dense (multitracer) survey, $M_{*,{\rm min}} = 10^{10} M_\odot$,
of $V \gtrsim 100 (h^{-1}$Gpc$)^{-3}$. With a sky coverage of $f_{\rm sky} = 0.75$,
the available volume out to $z = 1, 2, 3$ is $\approx 40, 170, 320 (h^{-1}$Gpc$)^{-3}$ respectively.
We thus confirm (see Section \ref{subsec:mmin}) that for a single-tracer type survey, we need
a survey with a very wide redshift range, whereas a dense, multi-tracer experiment could get sufficient information
at $z < 2$.

\subsection{Maximum wave vector and redshift accuracy}
\label{subsec:sigz}

\begin{figure*}[htbp!]
\includegraphics[width=0.48\textwidth]{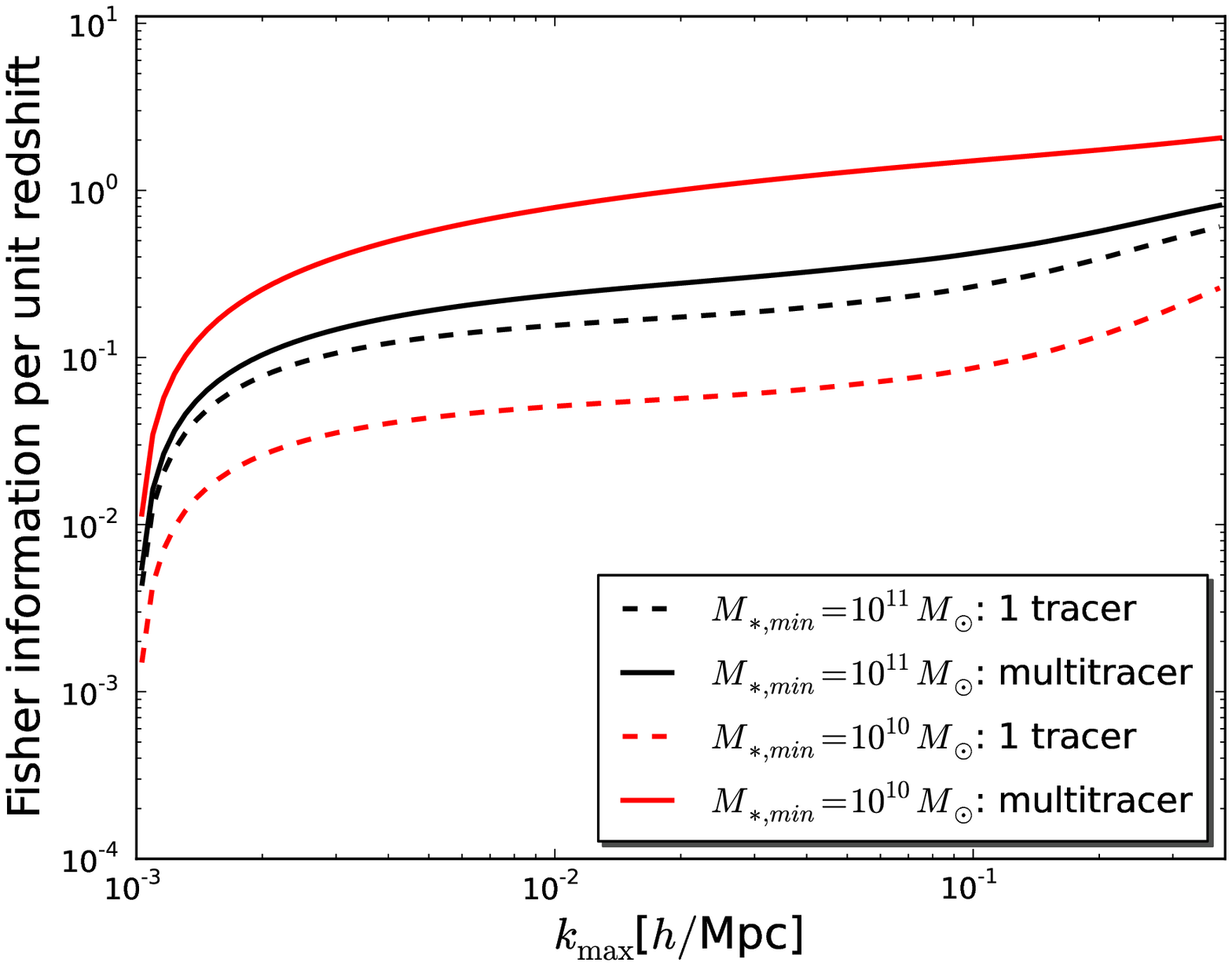} 
\includegraphics[width=0.48\textwidth]{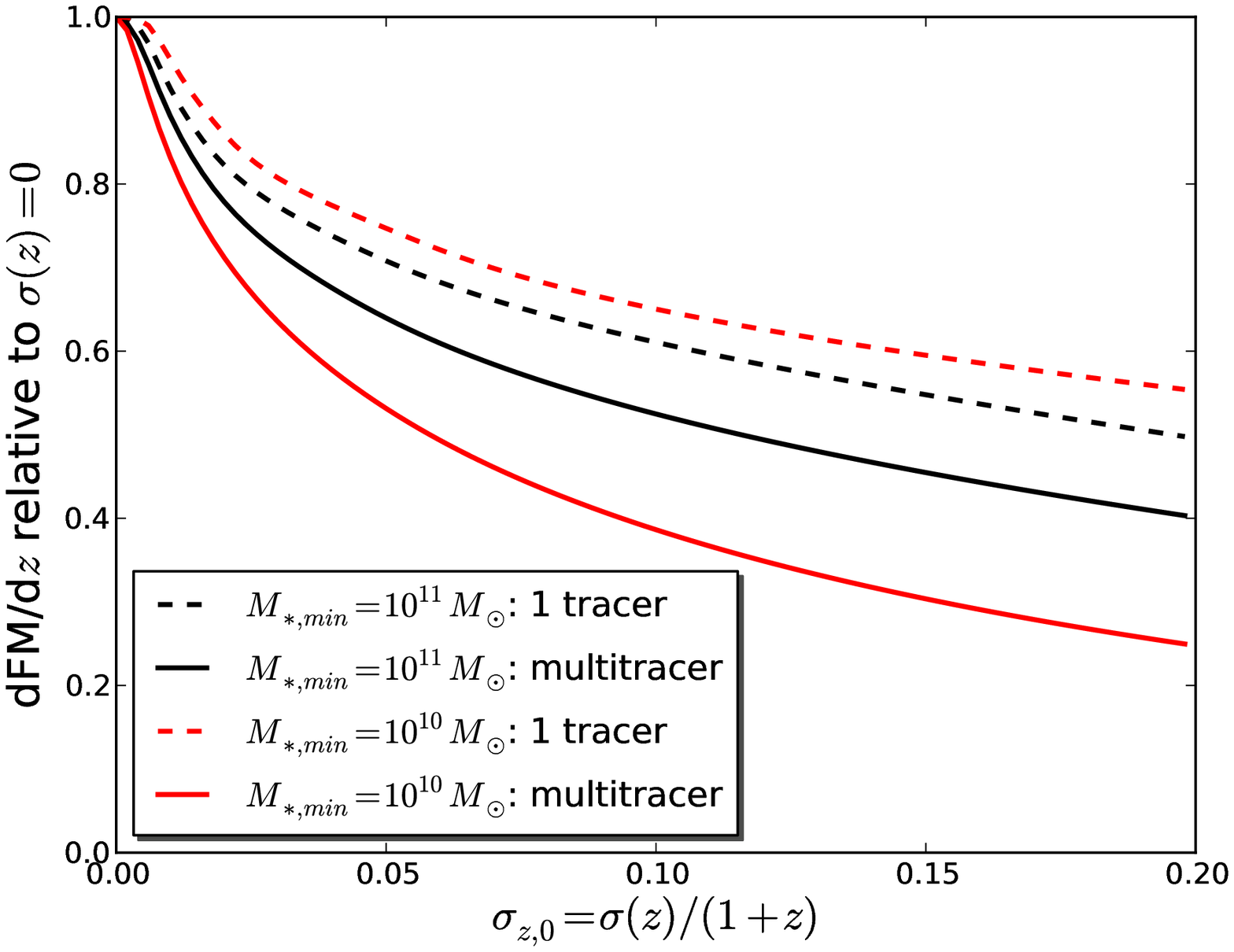} 
\caption{{\it Left:} Fisher information on $\fnl$ per unit redshift at $z = 1$ as a function of maximum
wave vector $k_{\rm max}$. We fix $k_{\rm min} = 10^{-3} h/$Mpc.
A large amount of the information on $f_{\rm NL}$ comes from the
very largest scales, $k < 0.01 - 0.02h/$Mpc. This relaxes the requirements
on redshift accuracy and modeling of nonlinearities in the galaxy power spectrum.
{\it Right:} The ratio of the Fisher information per unit redshift with a redshift uncertainty
$\sigma(z) = \sigma_{z,0} \, (1 + z)$ to the Fisher information with spectroscopic redshifts (modeled as $\sigma(z) = 0$).
We here assume the default $k_{\rm min} = 0.001 h/$Mpc, $k_{\rm max} = 0.1 h/$Mpc and $z=1$. The plot shows that even large redshift uncertainties,
$\sigma_{z,0} \lesssim 0.1$, are tolerable.
The multitracer constraints are a bit more sensitive to redshift accuracy as the information content on $\fnl$
is skewed toward smaller scales relative to the single-tracer case.}
\label{fig:FM vs kmax}
\end{figure*}

We now consider the dependence of $\fnl$ constraints on the smallest scales allowed in the analysis and
its consequences for the requirement on a survey's redshift accuracy.
The left panel of Figure \ref{fig:FM vs kmax} shows the
dependence of the Fisher information per unit redshift
on the smallest scales included in the Fisher forecast, $k_{\rm max}$.
We here fix the largest included scale to its default value, $k_{\rm min} = 0.001 h/$Mpc.
As seen in the figure, a large fraction of the constraining power comes from very large scales,
$k_{\rm max} \lesssim 0.01 - 0.02 h/$Mpc. Indeed, in the single-tracer case, and ignoring shot noise, the scaling with wave vector
of the Fisher information is
\be
F_{\fnl \fnl} \propto \int d\ln k \, k^{-1} \, T^{-2}(k).
\ee
At small $k$, the transfer function is constant, $T(k) \sim 1$,
so that $F_{\fnl \fnl}$ gets most information from $k$ closest to $k_{\rm min}$
and converges with increasing $k_{\rm max}$. At larger $k$, the transfer function term changes this picture.
For $k \gg k_{\rm eq}$ (the horizon scale at matter-radiation equality), $T \propto k^{-2} \ln(k/k_{\rm eq})$
so that more information is added at large $k$ and $F_{\fnl \fnl}$ in principle (i.e.~in the absence of
shot noise and ignoring nonlinearities) does not converge with $k_{\rm max}$.

In any case, the fact that so much information comes from $k \lesssim 0.01 - 0.02 h/$Mpc
is a good thing for at least two reasons. First of all, at $k \gtrsim 0.1 h/$Mpc,
non-linear effects on the matter density, galaxy bias and redshift space distortions
become increasingly difficult to model. We see that there is a lot of information on $\fnl$ that steers well
clear of this regime. Secondly, as we will see next, measuring the signal on these very large scales
does not require a high redshift accuracy.

A Gaussian scatter $\sigma(z)$ in a galaxy's estimated redshift relative to
its true redshift leads to an exponential damping in $k_z$, the line-of-sight
component of the wave vector, of the galaxy power
spectrum\footnote{Note that, contrary to the claim in \cite{asoreyetal12}, this damping can {\it not} be used as a signal
to constrain $H(z)$. The damping is simply given by $\sigma(z)$ divided by
whatever the value of $H(z)$ is in the fiducial cosmology that is used to convert redshift and angles
to positions. The damping thus does not tell us about the true value of $H(z)$.}
(but not in the shot noise), see e.g.~\cite{asoreyetal12},
\be
P(k,\mu) \to e^{-k^2 \mu^2 \sigma^2(z) c^2/H^2(z)} \times P(k,\mu).
\ee

Assuming a fiducial model for the redshift dependence of the redshift scatter,
$\sigma(z) = \sigma_{z,0} \, (1 + z)$,
the right panel of Figure \ref{fig:FM vs kmax} shows the degradation in Fisher information on $\fnl$
vs.~redshift scatter, relative to the case of perfectly known redshifts, $\sigma(z) = 0$.
For our purposes, the latter case corresponds to the level of redshift accuracy that would
be achieved with a spectroscopic survey.
The figure assumes the default wave vector range $k = 0.001 - 0.1 h/$Mpc.
We remind the reader that the degradation in the {\it uncertainty} on $\fnl$
scales like (the inverse of) the square root of the degradation in the Fisher matrix.
A ratio of a half in the figure thus implies an increase in uncertainty by $\sim 30 \%$.

The figure clearly demonstrates that the $\fnl$ constraining power is very robust against
redshift scatter. Uncertainties up to $\sigma_{z,0} \lesssim 0.1$ can be tolerated
without taking too big a hit on $\sigma(\fnl)$. This has important consequences for
the oprimization of
survey design, which we will discuss in the next Section.

The mild requirement on the redshift scatter for an $\fnl$-optimized galaxy survey
is reminiscent of the requirement on cosmic shear surveys.
However, while for such surveys a large redshift scatter is indeed acceptable,
the (photometric) redshift estimator's {\it distribution}, e.g.~its scatter and bias,
needs to be calibrated to sub-$\%$ level precision \cite{mahuhut06,hut02,hutetal04,hutetal06},
meaning that photo-$z$ calibration is potentially a major limiting systematic for upcoming cosmic shear
surveys.
This begs the question how well the redshift estimator needs to be calibrated for our type of survey.
Leaving a more thorough analysis for future work, as a first step to address this question, we perform a Fisher matrix forecast,
where, as before, we model the distribution of the redshift estimator $\hat{z}$
as a Gaussian,
\be
p(\hat{z}|z) \propto e^{-\frac{1}{2} \left( \hat{z} - z - \delta z \right)^2/\sigma_z^2},
\ee
where $z$ is the true redshift, but now we treat the scatter $\sigma_z$ as a free parameter and add an additional free parameter describing
a redshift offset, $\delta z$.
We then ask to what extent does a bias in $\sigma_z$ or $\delta z$ cause a bias in $\fnl$.
Since this is a question explicitly about parameter degeneracies, we include and marginalize
over other cosmological parameters: $\omega_b$, $\omega_c$, $\Omega_\Lambda$, $\sigma_8$ and $n_s$.

Our preliminary calculation assumes a single redshift bin centered at $z=0.5$ and uses only a single tracer.
We find that, while the redshift {\it offset} has a negligible effect on $\fnl$ for
reasonable values of $\delta z$, the {\it scatter} $\sigma_z$ needs to be calibrated to high precision.
Specifically, for a fiducial scatter based on the model $\sigma_{z,0} = 0.1$,
in order not to bias $\fnl$ by more than one standard deviation, $\sigma_{z,0}$
needs to be known to $\sigma(\sigma_{z,0}) = 0.014$ or better.
For a fiducial $\sigma_{z,0} = 0.01$, the scatter even needs to be calibrated
at the level $\sigma(\sigma_{z,0}) = 0.00053$.
Fortunately, if we fit the redshift scatter and offset to the data, simultaneously with $\fnl$
and other cosmological parameters, the redshift distribution parameters can easily be constrained to
the desired precision. In other words, they can be self-calibrated from the data (in weak lensing surveys, this may also
be possible if external spectroscopic data are available, see e.g.~\cite{phzpaper}).
However, self-calibration of course crucially relies on the model for $p(\hat{z}|z)$
being correct. While beyond the scope of this article, it will be important
to perform a more systematic study of the requirements on redshift calibration, that,
among other things, moves beyond the simple, Gaussian model considered here.

\subsection{Summary and discussion of survey optimization}
\label{subsec:sec summary}

We now summarize some of the main conclusions from the previous subsections
on what a galaxy survey aiming for $\sigma(\fnl) \sim 1$ should deliver.

\begin{itemize}
\item
The survey should cover a large volume, $V > 100 \, (h^{-1}$Gpc$)^{-3}$.
This is partially because for a fixed volume in ${\bf k}$-space the number of modes is
proportional to volume (and the number of modes is limited on large scales), but also because larger survey volumes
allow measurement of larger scales (smaller $k_{\rm min}$), which is where most of the scale-dependent bias
signal comes from.
\item
The survey does {\bf not} require very high redshift accuracy. Only when
$\sigma(z)/(1 + z) \gtrsim 0.1$, does most of the information on $\fnl$ get lost
due to smearing of the line-of-sight clustering signal.
\item
The survey needs a moderate to high depth galaxy sample, $M_{*,{\rm min}} \lesssim 2 \cdot 10^{11} M_\odot$.
Better constraints, or equivalently, the same constraints with a smaller volume,
can be obtained if multiple galaxy samples with different biases are used. For this multitracer approach, a very deep sample
is needed, $M_{*,{\rm min}} \lesssim 5 \cdot 10^{10} M_\odot$.
\item
The survey sample, and subsamples in the case where the multitracer technique is used,
need(s) to be based on cuts in an observable that strongly correlates with halo mass and therefore bias.
We have considered here stellar mass $M_*$ as the main observable, which in our default model has a scatter
$\sigma(\log_{10} M_*) \approx 0.25$ relative to the stellar mass to halo mass relation. Increasing this scatter, however,
to $\sigma(\log_{10} M_*) \sim 1$ would lead to up to an order of magnitude degradation in the signal-to-noise squared
of the $\fnl$ signal.

Since redshift errors propagate into the stellar mass uncertainty (see, e.g.,
Figure 4 of \cite{L12}), the requirement of a low-scatter measurement of stellar mass (or of another halo mass proxy)
in principle also places a requirement on the redshift scatter, in addition to the one discussed in the second bullet point.
We will not quantify this additional redshift requirement here, but stress that in principle it is contained
in any requirement on the accuracy of a measurement of stellar mass, intrinsic luminosity, etc.

\end{itemize}

Based on the above, we can now ask what type of galaxy survey is optimal for constraining $\fnl$.
Interestingly, an optimal $\fnl$ survey would look very different than the currently prevalent type of
cosmological galaxy clustering survey. Such surveys are typically optimized for baryon acoustic oscillations (BAO),
and other physics with a large signal down to small scales.
This naturally leads to surveys with {\it spectroscopic} redshifts (to measure
clustering down to $k_{\rm max} = 0.1 - 0.2 h/$Mpc) and galaxy number densities $\bar{n} P \sim 1$, where $P$ is the amplitude
of the galaxy power spectrum at some representative scale ($k \sim 0.1 - 0.2 h/$Mpc for BAO, but much smaller for
$\fnl$). Increasing the number density much beyond this does not
improve BAO constraints.
On the contrary, the mild redshift accuracy requirement and stringent requirements on survey volume
and number density strongly suggest that a spectroscopic survey is not optimal for $\fnl$, as a lot of time
would be spent achieving a better than needed redshift accuracy, limiting the total number of galaxies.

Instead, we argue that {\it a large area (ideally full-sky), multi-band, imaging survey would be ideally suited for constraining primordial non-Gaussianity
from scale-dependent halo bias.} The redshifts would thus be photometric redshifts, or, in the case
of a survey with a large number of narrow bands, low-resolution spectroscopic redshifts.
One can make use of {\it all} galaxies with a good enough redshift. Typically, this will mean a very high number density
at low redshift, with a decrease towards higher redshift. Thus, at low redshift, the multitracer technique can
be applied to boost the $\fnl$ constraint, whereas at the higher redshift end of the survey
using a single tracer is close to optimal and there is no multitracer benefit\footnote{An interesting question that we will come back to in
Section \ref{sec:survey} is which regime is more important to $\sigma(\fnl)$. In other words,
does the use of the multitracer technique at the low redshift end contribute strongly to the final $\fnl$ constraint
integrated over the redshift range of the entire survey?}.

To achieve $\sigma(\fnl) \sim 1$, the imaging survey needs to be rather deep, ideally achieving a high enough
number density, $\bar{n} > 10^{-4} (h^{-1}$Mpc$)^{-3}$, to at least $z \sim 2$, even for a full-sky survey.
Since photo-$z$ quality
redshifts are sufficient, this is not as stringent a requirement to fulfill as one might think based on intuition
developed from spectroscopic surveys.
Getting redshifts for a deep sample translates into requirements on the photometry,
i.e.~on the number of bands, their widths, the wavelength coverage, and the sensitivity per band.
Moreover, we need a wide enough wavelength coverage and high enough wavelength resolution to eliminate degeneracies
between SED templates that may lead to dangerous outliers in the distribution
of the redshift estimator.
Finally,
the requirement of good measurements of stellar mass, or another low-scatter proxy for halo mass,
also needs to be taken into account.
In practice, for instance, strong stellar mass measurements
will favor observing in the near infrared.
Since a large fraction of stellar mass information comes from the rest frame K band,
for a galaxy at redshift $z$, one would thus like to measure fluxes at wavelengths including
$\lambda \approx (1 + z) \, 2.2 \, \mu$m.
We will not attempt to further quantify the exact 
requirements on the photometry in this article, but simply note that the above considerations
would all need to be included in such a study.

\section{Putting it all together:\\including redshift dependence of the galaxy sample}
\label{sec:survey}

We have based our conclusions so far on a study at various redshifts of the $\fnl$ constraining
power as a function of minimum stellar mass, $M_{*,{\rm min}}$, and various other survey properties at
the given redshift. As mentioned above, a real multi-band imaging survey will have its
sample properties, such as $M_{*,{\rm min}}$, number density and bias, vary strongly with redshift.
We would thus like to model how such a survey constrains $\fnl$ {\it as a function
of redshift}. An accurate model for a realistic imaging survey would depend on many survey properties,
including the aforementioned photometry (sensitivity in each band, etc.), and would
also depend on currently poorly known properties of the galaxy populations that would be measured
by such a survey. Such a study is well beyond the scope of this article. Instead, in Section
\ref{subsec:toy}, we will consider a toy model for an imaging survey
to at least get an idea of how the sample properties may vary with redshift and how the $\fnl$
constraint depends on the total galaxy sample size/survey depth.
In Section \ref{subsec:real surveys}, we will then briefly comment on the prospects for
planned or proposed galaxy surveys.

\subsection{A toy model for an $\fnl$ galaxy survey}
\label{subsec:toy}

\begin{figure*}[htbp!] 
\includegraphics[width=0.48\textwidth]{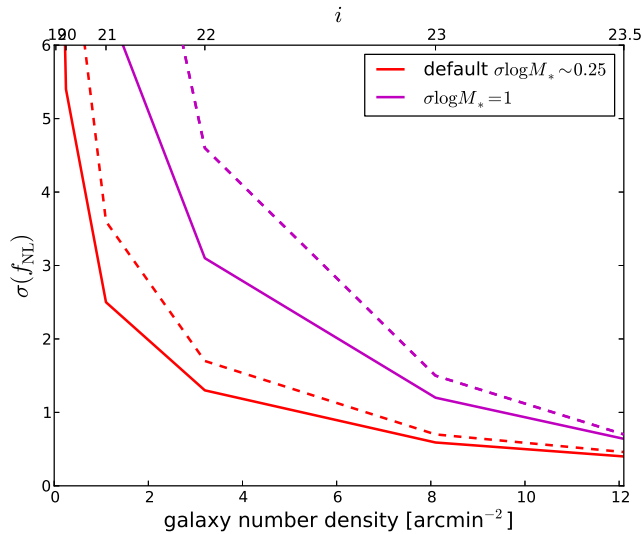}
\caption{Single-(dashed) and multi-(solid) tracer uncertainties on $f_{\rm NL}$ as a function
of an $i$-band AB magnitude cut. The cut specifies the number density as a function of redshift (estimated
from the COSMOS catalog).
Other properties of the sample, such as its galaxy bias, are computed using the toy model
described in the text. We assume a sky coverage $f_{\rm sky} = 0.75$.
We show results for two different values of the stellar mass scatter.
A full-sky survey of $\gtrsim 5$ galaxies arcmin$^{-2}$ could reach an order unity
$\fnl$ bound.}
\label{fig:sigvsz i cut}
\end{figure*}

The galaxy number density as a function of redshift
for our model survey is found by simply assuming
an $i$-band AB magnitude limited sample. We calculate this number density vs.~redshift directly
from the COSMOS catalog from \cite{tinkeretal13}, the same catalog
used to obtain the HOD parameters described in Section \ref{subsec:hod}.
Specifically, we use the Subaru $i^+$ filter \cite{capaketal07} to define the $i$-band magnitude cuts.
The COSMOS data have a depth $i^+ \sim 26.2$ (AB magnitude, $5\sigma$ in a $3''$ aperture).

To estimate the galaxy bias, and, in the multitracer case, the biases and number densities of the subsamples,
we again use the stellar mass based approach from the previous sections.
Specifically, at each redshift, we identify an effective $M_{*,{\rm min}}$
by matching the number density to the number density of the magnitude limited sample,
i.e.~we use the abundance matching technique.
We then calculate the bias at that redshift, and the number densities and biases of subsamples,
based on that $M_{*,{\rm min}}$,
as in the previous sections.
Throughout this section, we will assume a nearly full-sky survey, $f_{\rm sky} = 0.75$.

In this crude model, $i$-band magnitude is thus taken as an indicator for
detectability and for the redshift accuracy that could be achieved.
In a realistic scenario, these things would of course depend on a more complicated parameter space.
Moreover, we implicitly treat $i$ (or the galaxy properties it represents in our toy model)
as a proxy for stellar mass when we describe our sample in terms of $M_{*,{\rm min}}$.
Again, to model dispersion between the quantity that determines the sample selection and halo mass,
we consider two values of the scatter $\sigma_{\log M_*} = \sigma(\log_{10} M_*)$:
the scatter appropriate for stellar mass itself (as measured in the COSMOS sample
discussed previously), $\sigma_{\log M_*} \sim 0.25$, and a large scatter,
$\sigma_{\log M_*} = 1$.
We note that the actual relation between $i$ and $M_*$ at a given $z$ has a large scatter,
but that in reality we would be able to use a better quantity than $i$-band magnitude, like stellar mass itself
or estimated redshift accuracy,
to define sample cuts. The $i$-band cut is merely a simplified way of specifying the depth of the sample.

While this model is very simplistic, it will give us an idea of what a plausible redshift dependence
is for the galaxy sample properties of real multi-band imaging surveys of various depths.
The main results of this section are shown in Figure \ref{fig:sigvsz i cut}.
It depicts $\sigma(\fnl)$ as a function of the total number density of the survey (lower
horizontal axis) and the corresponding $i$-band limiting magnitude (upper horizontal axis).
As usual, solid curves employ the multitracer technique, dividing the sample into a large number of
subsamples and optimally combining them, while dashed curves use a single tracer\footnote{To not
undersell the single-tracer case, whenever we are in the regime where $M_{*,{\rm min}}$ is smaller
(i.e.~the sample deeper)
than the optimal value for $M_{*,{\rm min}}$ in the single-tracer case, we instead use that optimal value.
In other words, when the sample is so deep that the single-tracer case is weakened because of the bias approaching
unity, we assume we can throw away the low stellar mass part of the sample and apply the single-tracer
analysis to a more optimal subsample.}.
Different colors correspond to different scatters between halo mass and stellar mass (or the mass proxy it represents
in our toy model).

\begin{figure*}[htbp!] 
\includegraphics[width=0.48\textwidth]{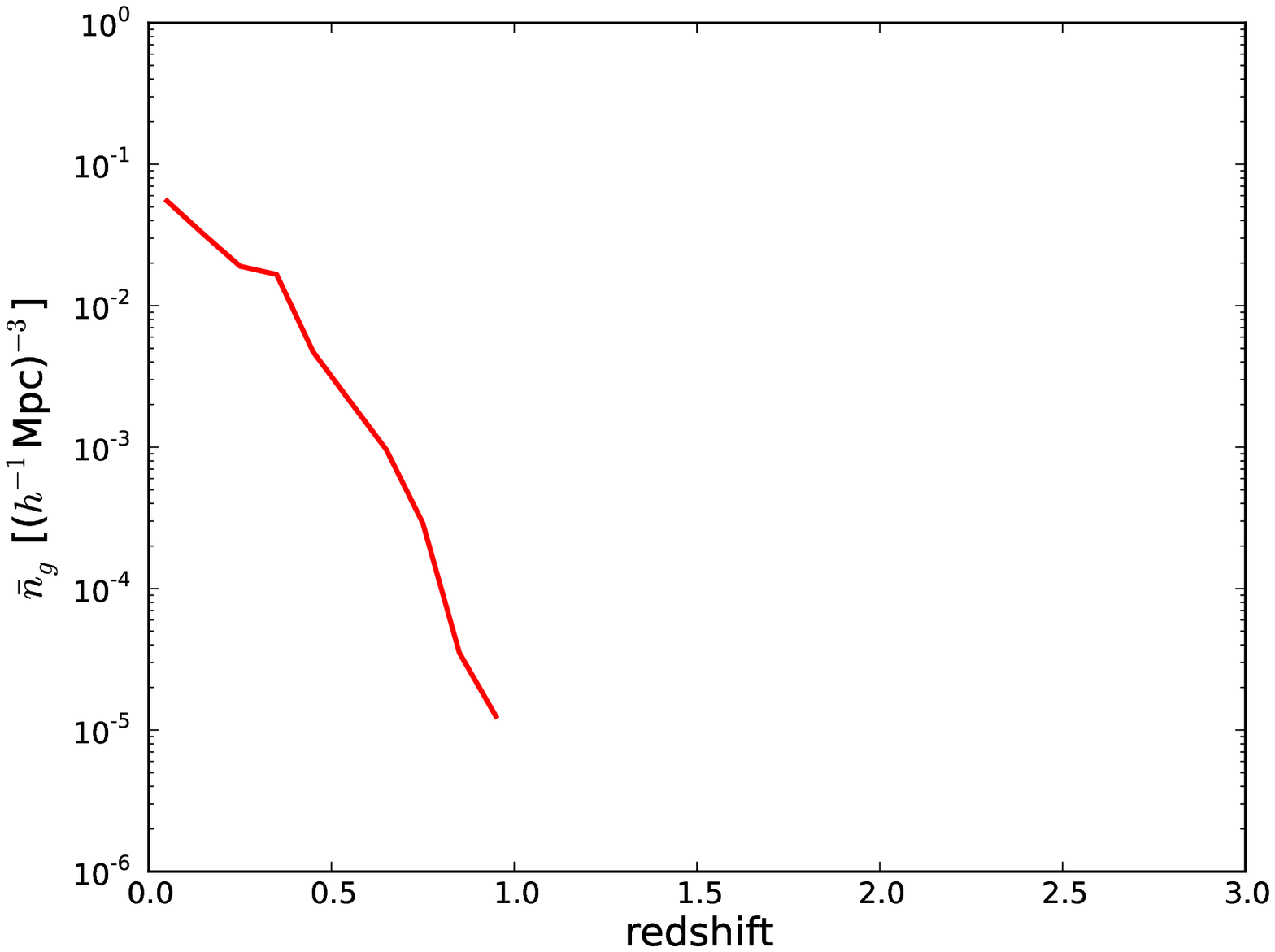}
\includegraphics[width=0.48\textwidth]{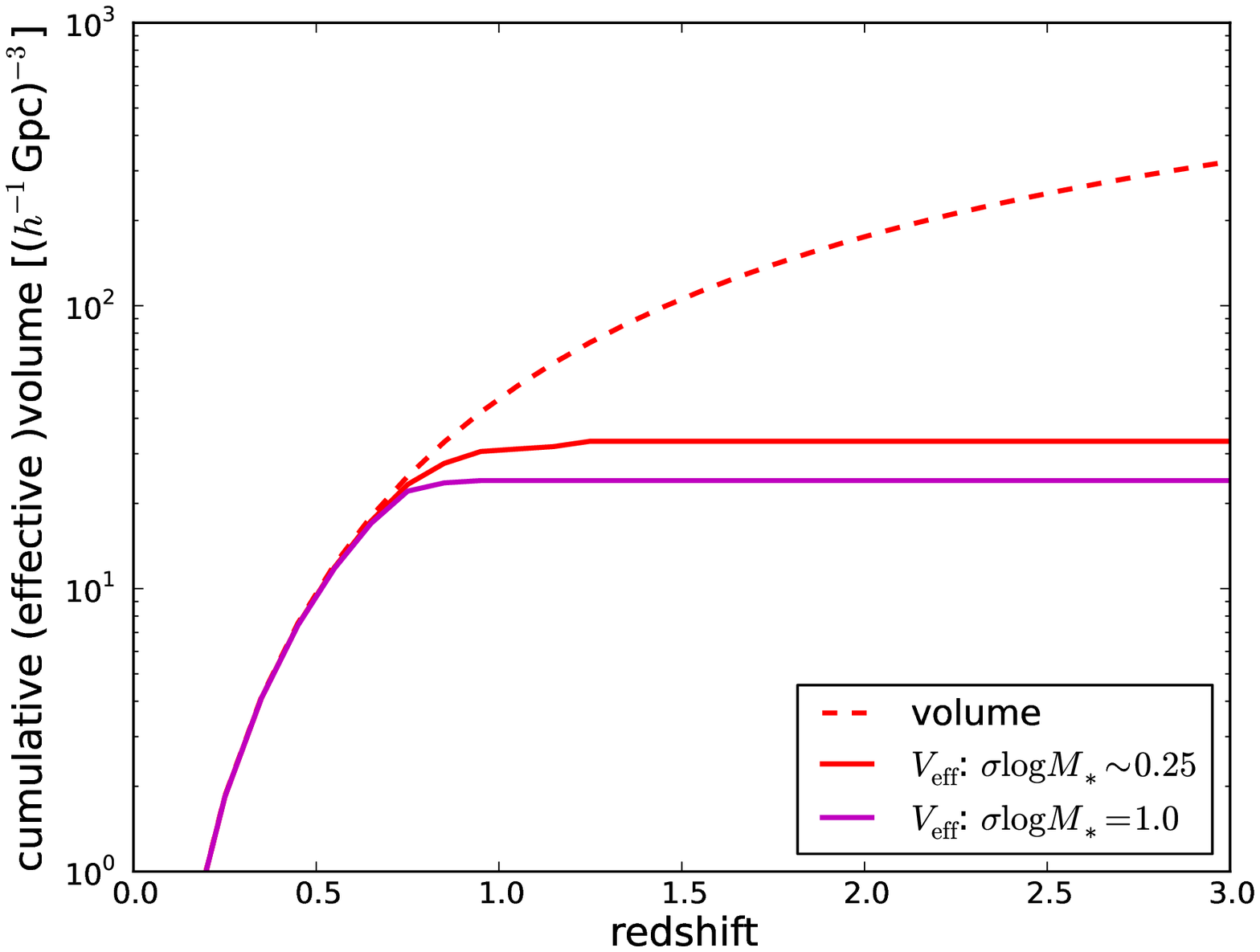}
\includegraphics[width=0.48\textwidth]{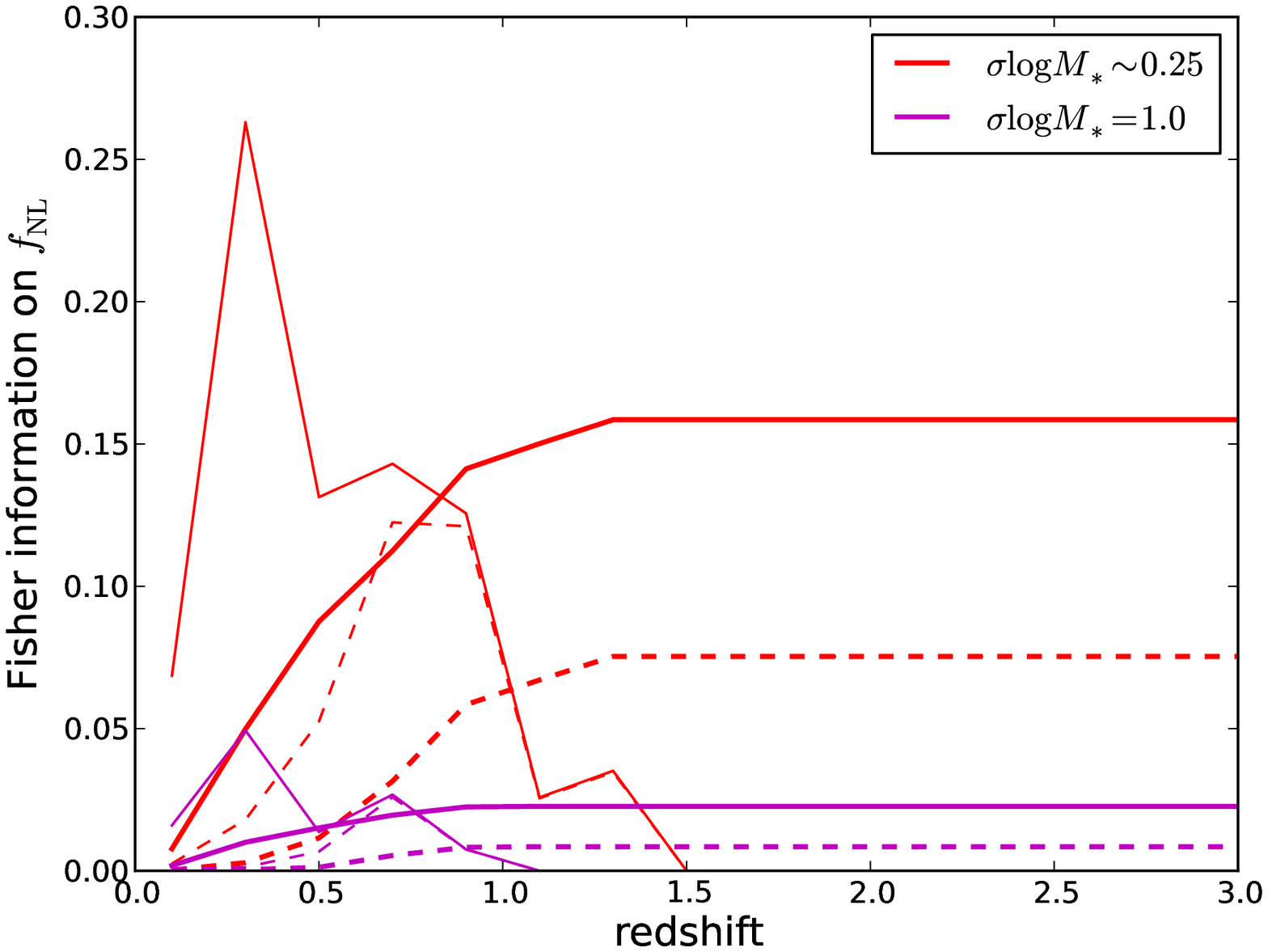}
\includegraphics[width=0.48\textwidth]{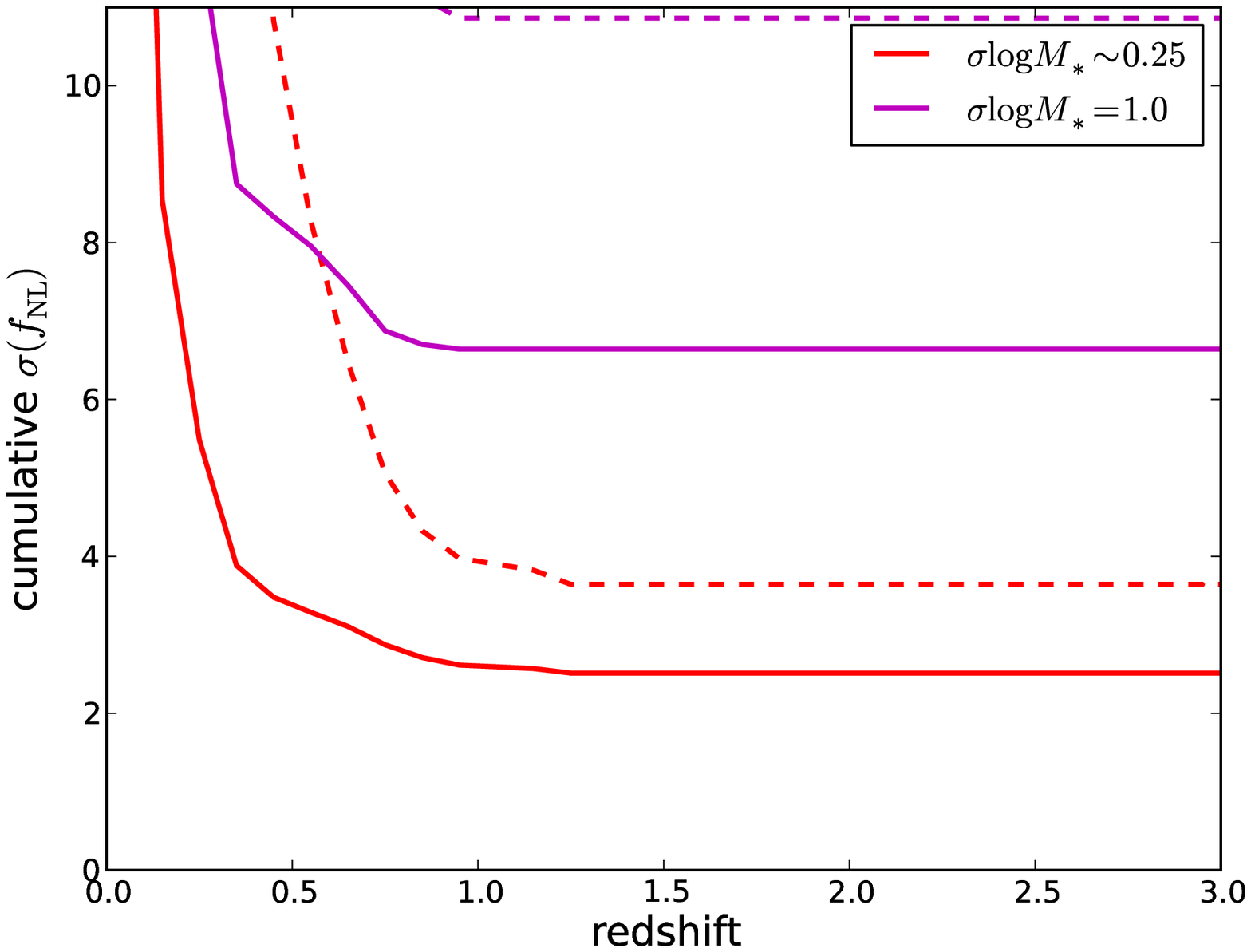}
\caption{The $f_{\rm NL}$ constraint as a function of redshift for a magnitude
limited sample in an imaging survey (see text for details) with $i < 21$ and $f_{\rm sky} = 0.75$.
{\it Top left:} Comoving number density. {\it Top right}: Cumulative volume and effective volume.
{\it Bottom left}: Cumulative Fisher information on $f_{\rm NL}$ (thick) and its derivative w.r.t.~redshift (thin).
Solid curves assume the use of multiple tracers and dashed curves assume a single tracer.
{\it Bottom right:} The uncertainty on $f_{\rm NL}$ based on all galaxies up to $z$. The
different colors indicate different scatters in the relation between stellar mass (which plays an important role
in our survey toy model) and halo mass.
There is clearly a low-redshift multitracer regime and a high-redshift single-tracer one. While the effect
of the multitracer technique on the final error bar on $\fnl$ is modest, this technique does allow a strong independent
constraint at $z < 1$, where there is not enough volume for the single-tracer technique to place a tight constraint.}
\label{fig:sigvszi21}
\end{figure*}

\begin{figure*}[htbp!]
\includegraphics[width=0.48\textwidth]{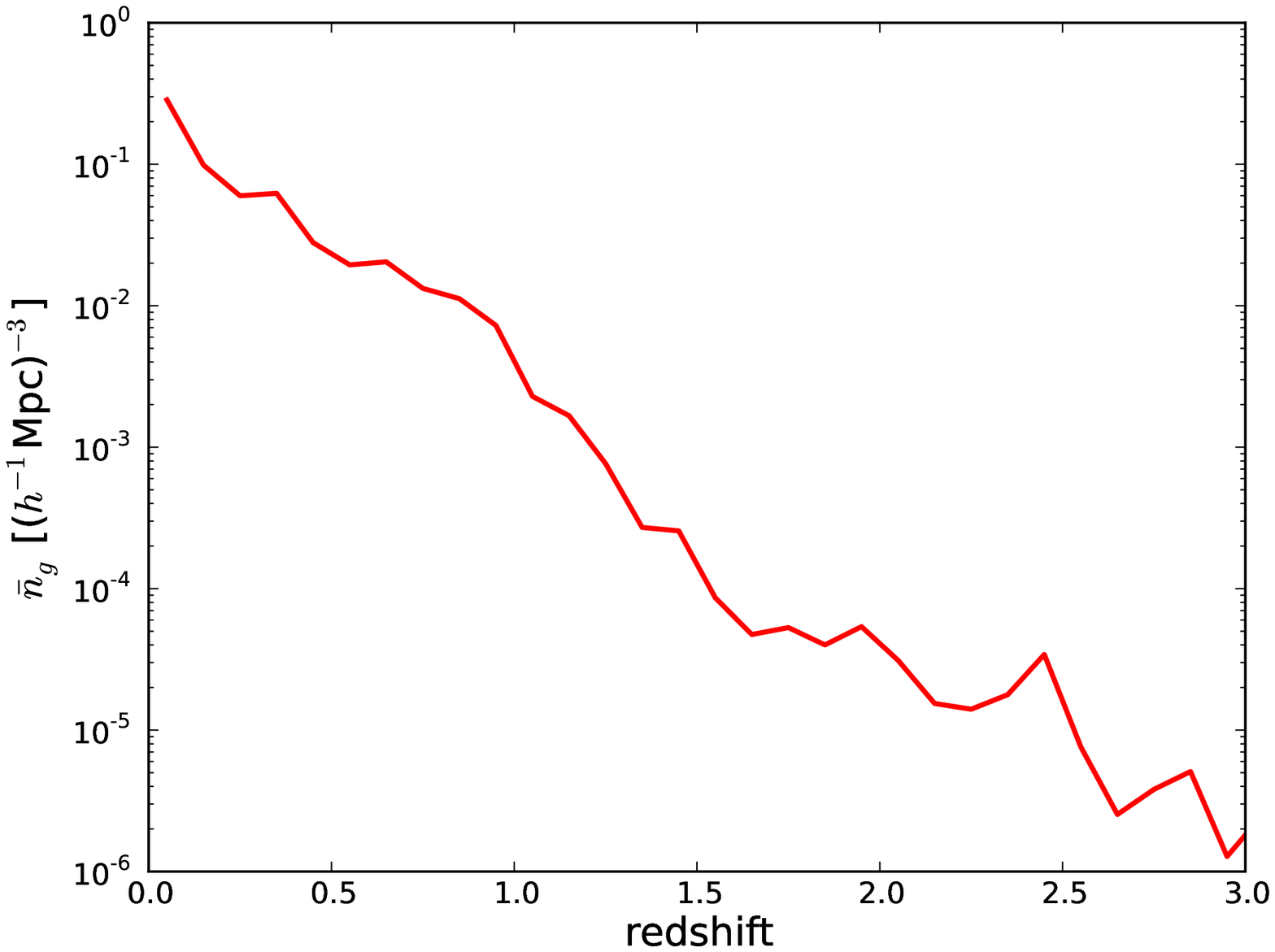}
\includegraphics[width=0.48\textwidth]{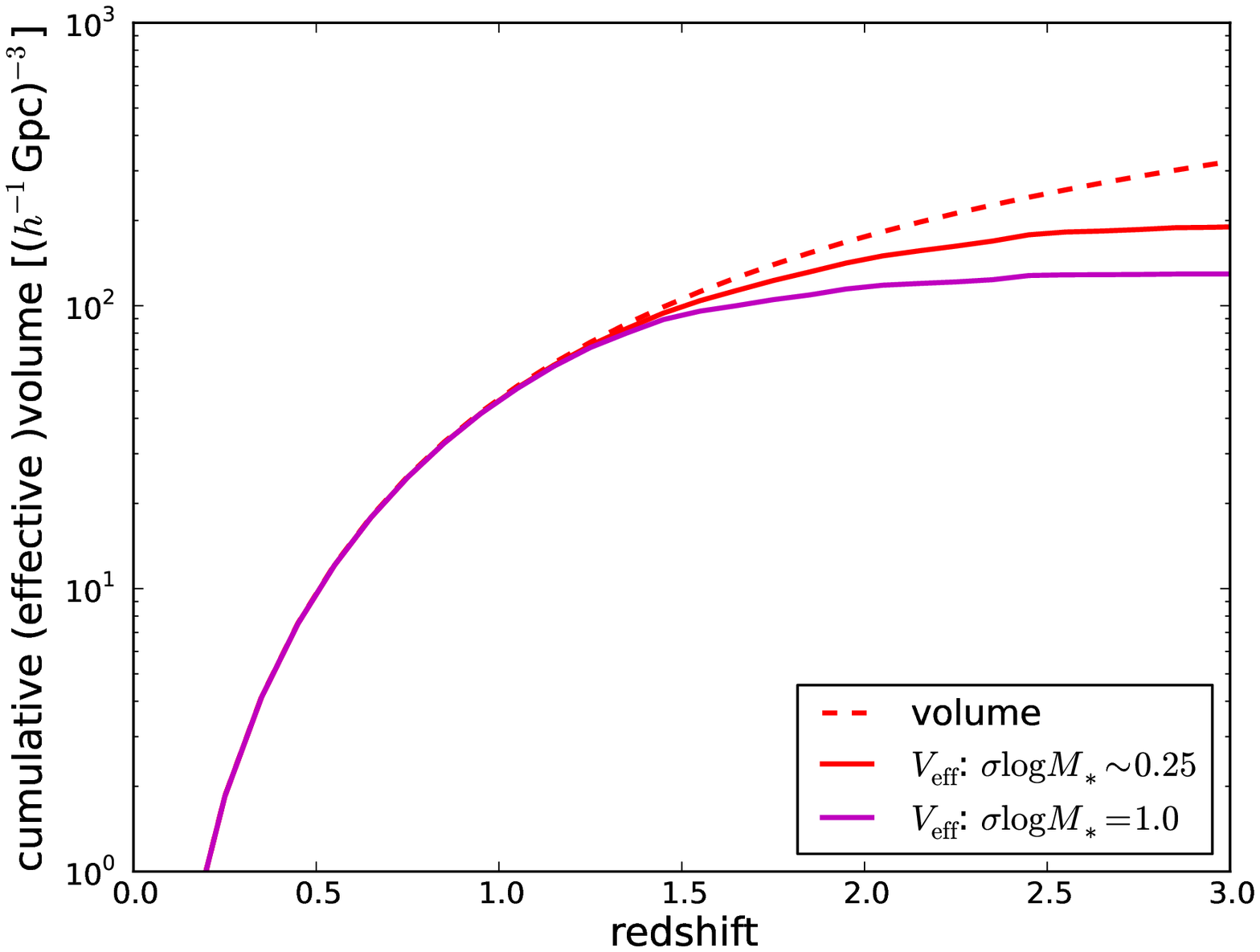}
\includegraphics[width=0.48\textwidth]{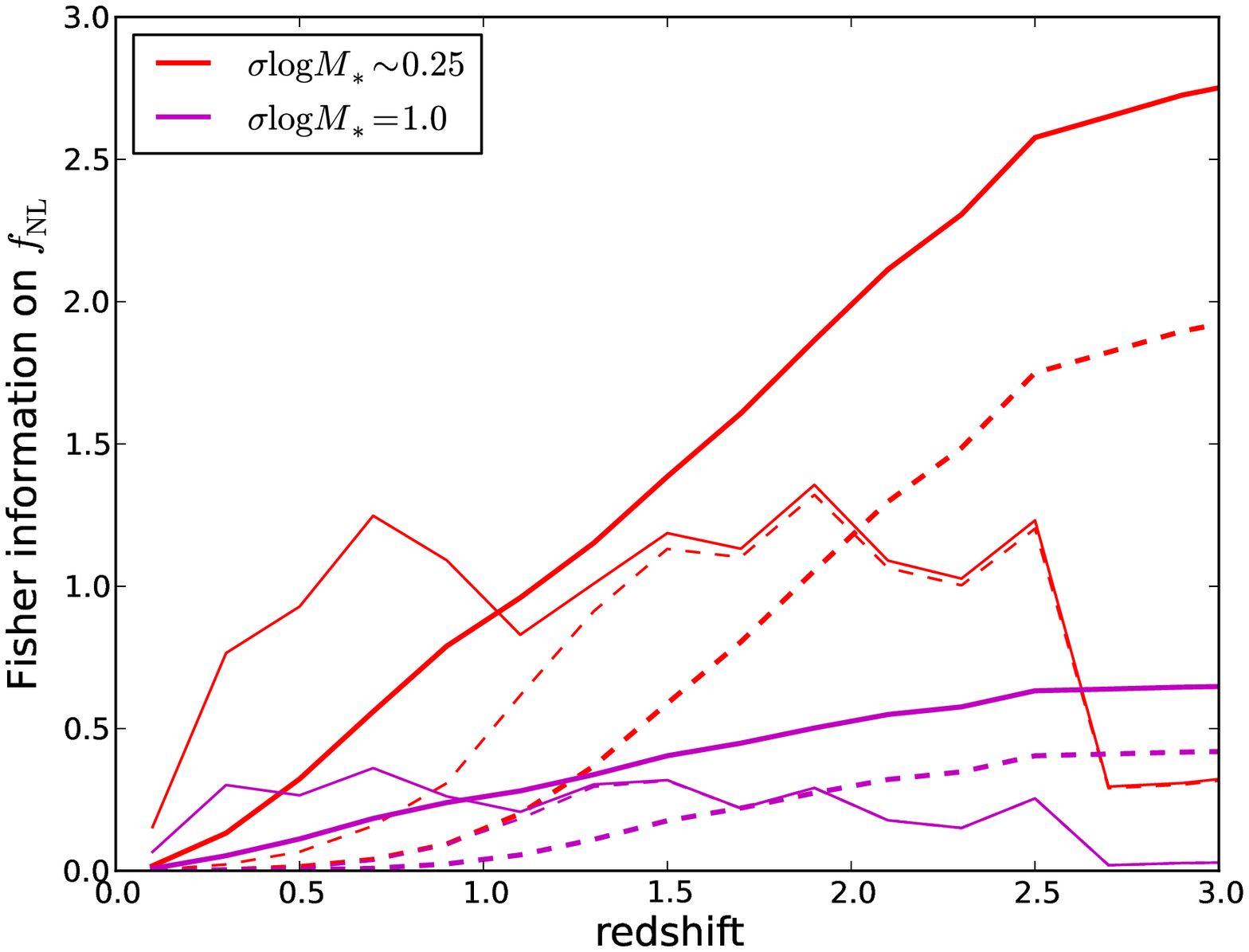}
\includegraphics[width=0.48\textwidth]{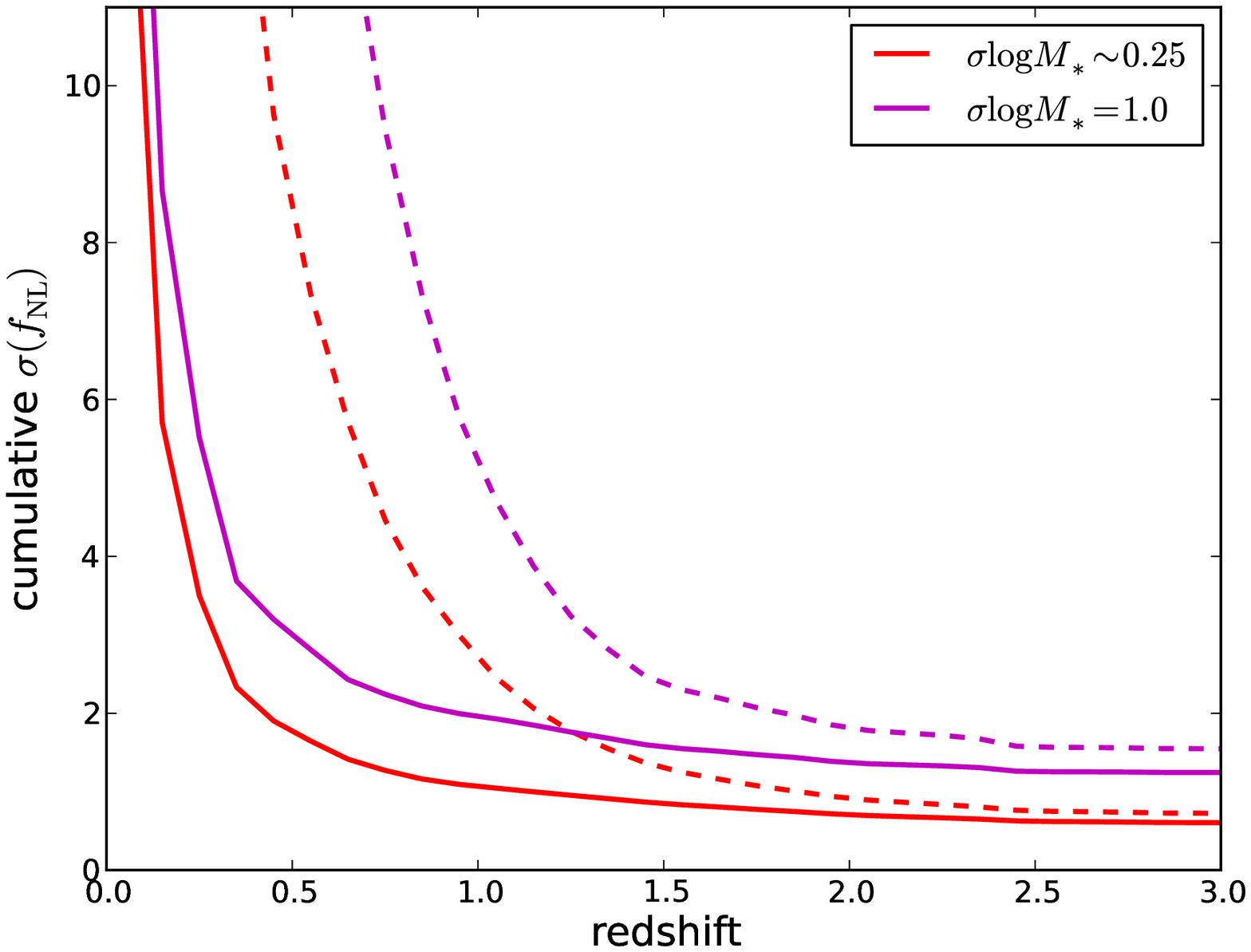}
\caption{As Figure \ref{fig:sigvszi21}, but for a deeper sample, $i < 23$.}
\label{fig:sigvszi23}
\end{figure*}

Focusing first on the case with default stellar mass scatter (red),
we find that order unity constraints on $\fnl$ can be achieved
for angular number densities $\bar{n}_A \gtrsim 5$ arcmin$^{-2}$.
In this regime, the multitracer technique leads to only a modest improvement
in the uncertainty. We will come back to the reason for this shortly.
Figure \ref{fig:sigvsz i cut} also demonstrates once again that having a large scatter
between the observed mass proxy (in this case stellar mass) and halo mass (and therefore halo bias)
is very detrimental, in this case causing a factor $\sim 2$ increase in $\sigma(\fnl)$.

To understand better how $\sigma(\fnl)$ in Figure \ref{fig:sigvsz i cut} builds up with redshift,
we focus on two examples in Figures \ref{fig:sigvszi21} and \ref{fig:sigvszi23}, corresponding
to $i < 21$ and $i < 23$ respectively. In each case,
the top left panel shows the comoving number density (again, obtained from the COSMOS
catalog) and the top right panel shows the resulting cumulative {\it effective}
volume\footnote{The effective volume depends on the power spectrum through $\bar{n}P(k)$, which, here,
is evaluated at a characteristic wave vector $k = 0.01 h/$Mpc.}
as a function of redshift.
The bottom left then shows the cumulative Fisher information on $\fnl$ (thick curves),
or, equivalently, the signal-to-noise squared of the $\fnl$ signal for $|\fnl| = 1$,
and the Fisher information per unit redshift, $d$FM$/dz$ (thin curves).
In the single-tracer case (dashed), one sees that $d$FM$/dz$ first grows with redshift
as the volume per unit redshift increases, then reaches a peak, and then starts to decline rapidly, as
the measurement becomes shot-noise dominated due to the declining galaxy number density.
The cumulative Fisher information reaches a plateau.
The difference in the multitracer case (solid) is that, at low $z$ and high number density,
it has a much larger $d$FM$/dz$ than the single-tracer case, but by the time the peak in the single-tracer
$d$FM$/dz$ curve is reached, the number density is low enough that the information from single- vs.~multitracer
have become equal.

For the importance of the multitracer approach to the final $\sigma(\fnl)$ value,
the question is thus which regime is more important? The low-redshift multitracer regime, or the high-redshift
single-tracer regime? The former has a higher signal per unit volume, but the latter may cover a larger volume.
In the case of our toy survey, the regimes turn out to be approximately equally important.
Therefore, the effect of the multitracer technique on $\sigma(\fnl)$, as shown in the right panel,
is not much more than a factor $\sim \sqrt{2}$ improvement. How this comparison works out for a realistic
survey depends on how steeply the comoving number density declines with redshift: a steeper function would increase the
importance of the multitracer regime relative to the single-tracer regime and vice versa (provided that at low redshift
the number density is high enough to benefit from multiple tracers).

While the multitracer gains may not look as impressive as expected when looking at $\sigma(\fnl)$ in the toy
model survey, an alternative way of looking at these results is that, thanks to the multitracer approach, one
can obtain two independent $\fnl$ measurements, each individually comparable to the constraint from single-tracer only
using the full survey. More concretely, for instance in the $i<23$ case,
one could get an order unity $\fnl$ measurement at $z <1$, heavily relying on the multitracer
approach, and an additional order unity measurement at $z > 1$, which does not benefit from multiple tracers
at all (but does from the large volume available at high $z$). Thus, the multitracer technique is a lot more
useful than suggested by the modest improvement in the value of $\sigma(\fnl)$ integrated over the entire survey.

\subsection{Upcoming/proposed surveys}
\label{subsec:real surveys}

We now relate the above to actual planned or proposed surveys.
As before, we will focus on the information that can be obtained from
the halo/galaxy {\it power spectrum} using the effect of scale-dependent bias.

Before discussing the forecasted numbers, first note that
in this paper we have restricted ourselves to the survey requirements
needed to reach the desired {\it statistical} error on $f_{\rm NL}$.
In addition, to obtain a believable constraint $\sigma(f_{\rm NL}) \sim 1$, any systematics
affecting the clustering measurement on the largest scales need to be controlled to high precision.
First of all, angular variations in, e.g., seeing (for a ground based survey),
stellar density, absorption by galactic dust or instrument sensitivity,
may lead to variations of the survey selection function (``depth'')
on large scales,
which, if not modeled, will lead to spurious clustering.
Secondly, insufficiently modeled redshift errors would modify the clustering
signal and could bias the determination of $f_{\rm NL}$.
A signal from $f_{\rm NL} \sim 1$ corresponds to variations of the relative galaxy overdensity
$\delta_g \sim 3 \cdot 10^{-4}$ on the largest scales of the survey, $10 - 100$ degrees,
where most of the constraining power comes from.
While it is beyond the scope of this paper to quantify the resulting requirements on
survey design, we note that there is thus an extremely stringent requirement
on systematics control.

We now turn to constraints from real world surveys.
First of all, the tightest current constraints are driven by clustering of 
galaxies and quasars in the Baryon Oscillation Spectroscopic Survey (BOSS),
and have error bars $\sigma(f_{\rm NL}) \sim 20$ \cite{rossetal13,giannantonioetal14,leistedtetal14}.
Systematics are a significant part of the error budget.
While current constraints are thus not competitive with the CMB
($\sigma(f_{\rm NL}) \approx 6$ from the Planck temperature bispectrum),
near-future large-scale structure surveys are expected to reach comparable or better constraints.

Considering first spectroscopic surveys of galaxies (and quasars),
the largest volumes will be probed by EUCLID \cite{euclid}
and DESI \cite{DESIwhitepaper}, which both approach $V \sim 100 (h^{-1}$Gpc$)^{-3}$, but with number densities
too low for the multitracer technique to lead to large gains.
The $f_{\rm NL}$ constraints, based on the power spectrum,
from these surveys are projected to be comparable to
the current CMB bound, e.g.~\cite{fontetal13} finds $\sigma(f_{\rm NL}) = 3.8$ for DESI and
$6.7$ for EUCLID (combined with BOSS).

In addition to spectroscopic surveys, there are many planned and ongoing cosmological imaging surveys.
Often with cosmic shear the main cosmology target, these surveys typically use a handful
of passbands to obtain photometric redshifts with an expected accuracy of order
$\sigma(z) \approx 0.05 (1 + z)$. Examples of such surveys that will probe the largest volumes
are the EUCLID imaging survey and LSST.
In terms of volume and number density, these surveys meet the requirements to reach $\sigma(f_{\rm NL}) \sim 1$
set in the previous sections.
For example, EUCLID expects to reach a number density of $\bar{n}_A \approx 30$ arcmin$^{-2}$
for their photometric redshift sample, corresponding to a magnitude 24.5 completeness cut
in their optical band. Comparing this to Figure \ref{fig:sigvszi23}, which shows our toy model
case $i < 23$, with $\bar{n}_A \approx 8$ arcmin$^{-2}$,
suggests that, even though EUCLID is not a full-sky survey (Area $\approx 15,000$ deg$^{-2}$),
its sample is so deep that it should be very competitive for $f_{\rm NL}$.
However, with only a small number of passbands, it will be particularly challenging to
reach the desired redshift calibration, to control the systematics discussed above
to high enough precision, and to obtain a low-scatter halo mass proxy for each galaxy (needed
to divide the sample into subsamples).
If these issues can be dealt with, our work implies that imaging surveys such as EUCLID and LSST
can in principle reach constraints $\sigma(f_{\rm NL}) \approx 1$ (see also, e.g., \cite{yamauchietal14}).

The potential challenges for a photometric redshift survey mentioned above may be more easily
addressed with an approach somewhere
in between those of spectroscopic and standard photometric surveys.
Specifically, an imaging survey in dozens of narrow bands will enable quasi-spectroscopic
redshift quality.
Indeed, the proposed satellite mission SPHEREx \cite{SPHEREx} is exactly such a survey and
has the measurement of $f_{\rm NL}$ to order unity precision as one of its main science goals.
This survey would measure a low-resolution spectrum across the full sky in $\sim 100$ bands
in the near infrared, $\lambda \approx 0.5 - 5 \mu m$. These data would allow the measurement
of redshifts with accuracy $\sigma(z) < 0.1 (1 + z)$ for $\sim 340 M$ galaxies (and much lower redshift uncertainty
for a subsample), covering
the full-sky from redshift zero to $z > 2$ with sufficient number density.
Indeed, a power spectrum analysis of this sample
would lead to $\sigma(f_{\rm NL}) \approx 1$.
Another narrow-band imaging survey, but in the optical and from the ground ($8,000$ deg$^2$), is J-PAS \cite{jpas},
which also expects to measure $\sigma(f_{\rm NL}) \sim 1$.

Beyond optical and infrared galaxy surveys, in principle the largest volumes can be probed
at radio wavelengths using the 21cm line emitted (or absorbed at certain redshifts) by neutral hydrogen
around the reionization epoch ($z \sim 10$). A future radio survey like SKA may
in principle measure $\sigma(f_{\rm NL}) \sim 1$ \cite{yamauchietal14,cameraetal14}, provided that foregrounds, which are four orders
of magnitude above the signal, can be subtracted out
to high precision.

Finally, while we have here focused on the information in the power spectrum,
the bispectrum also contains a signal from scale-dependent bias, in addition to a signal
from the primordial matter bispectrum itself, and is another excellent
probe of primordial non-Gaussianity. The bispectrum can be measured from any survey from which
the power spectrum can be measured and is in that sense an additional signal, that ``comes for free''.
Forecasts suggest that $f_{\rm NL}$ constraints from the bispectrum are better than those
from the power spectrum by at least a factor of two. In addition, the bispectrum constraint relies
on smaller scales than the power spectrum constraint and can therefore be seen, to an extent,
as an independent probe.

\section{Conclusions}
\label{sec:conclusions}

We have studied the ability of galaxy clustering surveys
to constrain primordial non-Gaussianity of the local type
to a precision beyond what is possible with the CMB.
We have set as a specific target an order unity constraint,
$\sigma(f_{\rm NL}^{\rm loc}) \sim 1$
(we typically drop the superscript).
This is motivated by the fact that, without fine-tuning, multi-field inflation
models generically predict $|f_{\rm NL}| \gtrsim 1$, and by the desire to improve relative to existing
and ideal future CMB bounds ($\sigma(f_{\rm NL}) \approx 6$ and
$\sigma(f_{\rm NL}) \approx 2$ respectively).
We have focused on the signal from scale-dependent halo bias as manifested
in the galaxy/halo power spectrum, leaving a bispectrum analysis for future work,
and considered both constraints from a single tracer and from a multitracer analysis.

In Section \ref{sec:survey optimization}, we have studied the dependence of the expected
$f_{\rm NL}$ bound on various survey properties (see Section \ref{subsec:sec summary}
for a detailed summary). A key consideration in this optimization study is the fact that the information
on $f_{\rm NL}$ is dominated by the largest scales accessible to the survey.
We concluded that to reach $\sigma(f_{\rm NL}) \sim 1$ with a {\it single} sample of galaxies
(or of another tracer of the underlying matter distribution),
comoving number densities of a few $\times \, 10^{-4} (h^{-1}$Mpc$)^{-3}$ are required,
corresponding to $\bar{n} P(k_c) \sim 1$ where $k_c$ is the characteristic scale providing information
on $f_{\rm NL}$, and $P(k)$ is the tracer power spectrum. This requirement is similar to,
and even slightly looser than, the number density requirement for a BAO survey.
Moreover, a very large survey volume, of at least a few $\times 100 \, (h^{-1} $Gpc$)^{3}$
and a redshift accuracy of $\sigma(z) \lesssim 0.1 (1 + z)$ are required.
To take advantage of the multitracer technique, a much larger number density is needed,
$\bar{n} P(k_c) \gg 1$ (in practice $\bar{n} \gtrsim$ few $\times \, 10^{-3} (h^{-1}$Mpc$)^{-3}$),
in which case the volume requirement is loosened to $V \gtrsim 100 (h^{-1} $Gpc$)^3$.
In general, one also needs to be able to measure a proxy for host halo mass with a relatively low scatter,
such as stellar mass of the central galaxy,
to select an optimal clustering sample or subsamples.

Looking at the upcoming spectroscopic galaxy surveys that probe the largest volumes,
experiments such as EUCLID and DESI approach $V = 100 \, (h^{-1}$Gpc$)^3$,
with number densities for which the multitracer method does not lead to large
improvement in the constraining power.
As a consequence, while not reaching $\sigma(f_{\rm NL}) \sim 1$,
these surveys are expected to obtain constraints competitive with those from the CMB.

Partially based on the loose redshift accuracy requirement,
we concluded that an imaging survey with photometric or low-resolution spectroscopic
(in the case of a narrow-band imaging survey) redshifts may be ideally suited to constrain
primordial non-Gaussianity. Such surveys can in principle
probe very large volumes more easily than spectroscopic surveys, at the
(acceptable) cost of lower redshift accuracy.
In Section \ref{sec:survey},
we used a toy model for such an imaging survey
to study the constraining power as a function of redshift
and the total constraint as a function of survey depth or total number density.
We found that a full-sky survey complete to magnitude $i \sim 23$,
corresponding to $\bar{n}_A \approx 8$ arcmin$^{-2}$,
should be able to reach $\sigma(f_{\rm NL}) = 1$.

Anticipating real world imaging surveys,
planned photo-$z$ surveys such as EUCLID and LSST
are expected to obtain significantly deeper samples
than what was discussed above and, despite their sky coverage being about half the full sky,
would thus probe the required, large volumes.
On the other hand, due to the limited number of wavelength bands of these experiments
redshift calibration and other systematics are a particularly serious concern.
Better redshift information can in principle be extracted with a narrow band,
high-resolution photometric survey, like SPHEREx or J-PAS. In particular,
SPHEREx is a proposed full-sky survey that will measure a spectrum using $\sim 100$ bands
in the near infrared, and one of its explicit goals is to reach $\sigma(f_{\rm NL}) \sim 1$.

Topics that require further study include large-scale systematics and how to control them
at the level set by $\sigma(f_{\rm NL}) \sim 1$, the constraining power of the bispectrum (which supersedes that
of the power spectrum in preliminary studies), and constraints on equilateral and other types of non-Gaussianity.


\acknowledgments

We thank Alexie Leauthaud and Peter Capak for
sharing their expertise on the observational properties of galaxies.
In addition, we thank Alexie Leauthaud for providing
the COSMOS catalog used in our Section \ref{sec:survey}.
Part of the research described in this paper was carried out at the 
Jet Propulsion Laboratory, California Institute of Technology, 
under a contract with the National Aeronautics and Space 
Administration. This work is supported by NASA ATP grant 11-ATP-090.

\appendix

\section{Stochastic bias beyond Poisson noise}
\label{sec:nonpoisson}

\begin{figure*}[htbp!] 
\includegraphics[width=0.48\textwidth]{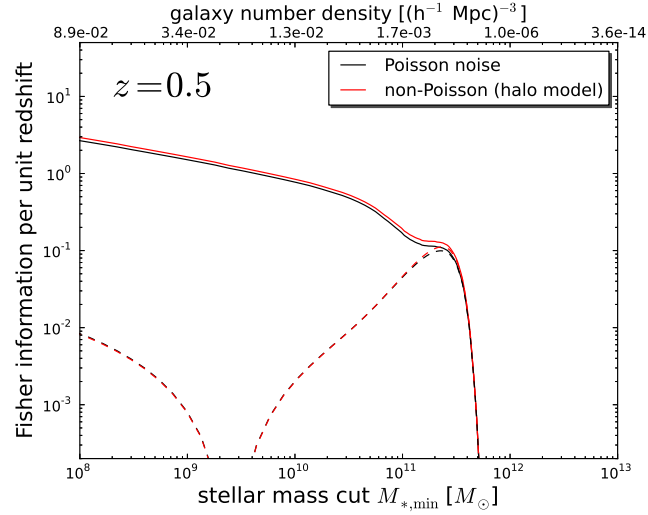}
\caption{The Fisher information on $\fnl$ per unit redshift (cf.~Figure \ref{fig:FMgal}), comparing
two prescriptions for the stochastic noise component in the galaxy clustering measurement.
The black curves correspond to our default approach of Poissonian shot noise given by the inverse
number density. The red curves use the halo model prescription from \cite{hamausetal10}.
For simplicity, we only show the case of zero stellar mass scatter relative to the mean stellar mass - halo mass
relation (in other words, cuts are essentially based on halo mass itself).
The difference in results between the stochastic noise prescriptions is not very large, $\leq 15 \%$ at
$z = 0.5$, decreasing toward higher redshifts.}
\label{fig:nonpoisson}
\end{figure*}

In the body of this paper,
we have assumed that the stochastic part of the halo clustering (as traced by galaxies)
can be approximately described by the second term on the right hand side of Eq.~(\ref{eq:pk kaiser}),
i.e.~diagonal shot noise given by one over the number density.
Using simulations, \cite{hamausetal10} have found deviations
from this description: the shot noise of the highest mass halos is actually
less than $1/\bar{n}_i$ and there are small off-diagonal correlations
between different halo bins. In the same work, it was also found that a more accurate
description of the stochastic noise is given by the halo model (see Eq.~(32) in \cite{hamausetal10}).

We have compared the $\fnl$ constraints expected in our default Poissonian
model to those using the halo model and found that, given the precision aimed at
in this paper, the Poissonian approximation is good enough. As illustrated in
Figure \ref{fig:nonpoisson} for the case of zero scatter in the halo mass proxy (stellar mass in this case),
the Poissonian approximation underestimates the Fisher information by at most $15 \%$ at $z = 0.5$, and by less
at higher redshift.

\section{Multi-tracer constraints as a function of number of mass bins}
\label{sec:mass bins}

\begin{figure*}[htbp!] 
\includegraphics[width=0.48\textwidth]{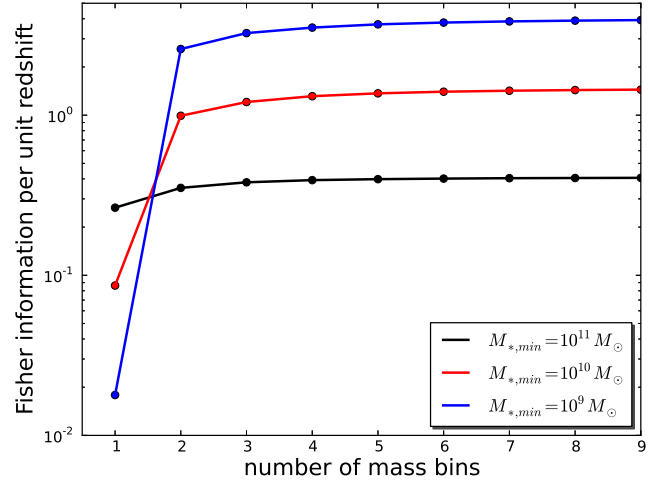}
\caption{Fisher information on $f_{\rm NL}$ per unit redshift (at redshift $z = 1$)
as a function of the number of stellar mass bins used for the  multitracer analysis.
Stellar mass bins are logarithmically spaced between $M_{*,{\rm min}}$
and $M_{*,{\rm max}} = 10^{12} M_\odot$. Three different survey depths are shown. The deeper the survey (lower $M_{*,{\rm min}}$),
the stronger the benefits of the multitracer approach.
It is clear that no more than three bins are needed in practice to take near-optimal advantage of the multitracer
information.
}
\label{fig:FM vs nbins}
\end{figure*}

Whenever we present multitracer constraints in this article, we use a number of stellar mass bins
large enough for the constraining power to converge so that our results represent the best possible
multitracer constraints. Considering how many bins are needed in practice to achieve these optimal constraints,
we find that typically (at most) three bins is sufficient. We illustrate this in Figure \ref{fig:FM vs nbins},
where we show the Fisher information per unit redshift at $z = 1$, for the case with default stellar mass scatter.
We divide the sample into subbins logarithmically spaced in stellar mass between $M_{*,{\rm min}}$ and
$M_{*,{\rm max}} = 10^{12} M_\odot$, and study three values of $M_{*,{\rm min}}$, spanning the range from a moderate-density,
single-tracer survey ($M_{*,{\rm min}} = 10^{11} M_\odot$) to a ultra-high-density, multitracer survey ($M_{*,{\rm min}} = 10^{9} M_\odot$).
The figure shows that using multiple bins is most important for the deeper samples, but that, even there, no more than three bins are needed
to reach optimal constraints.

\section{Marginalization over cosmological and nuisance parameters}
\label{sec:marg}

In our forecasts in this paper, we have ignored any degeneracies between $\fnl$ and
other cosmological and/or nuisance parameters. In general,
such degeneracies can have a huge effect on parameter constraints, weakening the
uncertainties on certain parameters by orders of magnitude relative to the unmarginalized
case. Fortunately, the effect of marginalization on $f_{\rm NL}$ is much more modest,
due to its unique, large-scale signature that is hard to mimic with other parameter combinations.

In a forecast such as ours, marginalization can be easily taken into account by computing the full Fisher matrix
including all relevant parameters and inverting it. To quantify the size of the effect, we have done such a calculation
for a survey of volume $V = 100 \, (h^{-1}$Gpc$)^3$ centered at z=1, considering the minimal cosmological
parameter space spanned by $\omega_b, \omega_c, \Omega_\Lambda, \sigma_8, n_s$ and a free linear, Gaussian bias parameter $b_i$
for each galaxy sample. We have considered both a single-tracer survey ($M_{*,{\rm min}} = 10^{11} M_\odot$)
and a multitracer survey with three stellar mass bins, logarithmically spaced in $M_*$ between
$M_{*,{\rm min}} = 10^{10} M_\odot$ and $M_{*,{\rm max}} = 10^{12} M_\odot$. Before (after) marginalization,
we find $\sigma(\fnl) = 1.71 (2.19)$ for the single-tracer scenario, and $\sigma(\fnl) = 0.86 (1.05)$
for the multitracer survey.

The effect of marginalization is thus typically a degradation in $\sigma(\fnl)$ of order $20 - 30 \%$
and is less strong in the multitracer case (since there, to an extent, the bias is measured directly
and no other parameters cause a scale-dependent bias).
The effect is thus not unimportant and should certainly be taken into account
for any concrete $\fnl$ forecast for specific surveys. However, the effect is small
enough that, for our goals of quantifying what survey is needed to obtain $\sigma(\fnl) \sim 1$
and of understanding the various trends with survey properties, it is justified to ignore it.

\bibliography{refs}

\end{document}